\documentclass{imamat}

\usepackage{latexsym,amssymb,lastpage}
\usepackage{graphicx,amsfonts,float,placeins}
\usepackage{times,mathptmx,bm,amsmath}
\usepackage{dcolumn}
\usepackage{mathrsfs}
\usepackage{tikz}
\usepackage{soul,color}
\usepackage[normalem]{ulem}

\usetikzlibrary{decorations.markings}
\usetikzlibrary{decorations.pathmorphing}
\usetikzlibrary{fit}
\usetikzlibrary{calc,arrows,patterns}
\usepgflibrary{plotmarks}

\usepackage{paralist}

\newcommand{\tr}{\mathrm{tr}\,}

\newcommand{\M}{\mathrm{max}}

\newcommand{\ti}{\mathrm{ti}}
\newcommand{\iso}{\mathrm{i}}
\newcommand{\T}{\mathrm{T}}

\newcommand{\tanhs}{\mathrm{\tanh\,}}
\newcommand{\C}{\mathrm{cyc}}

\numberwithin{equation}{section}

\begin{document}

\title{Multicyclic modelling of softening in biological tissue}
\author{{\sc Stephen R. Rickaby}\thanks{Email: stephen.r.rickaby@gmail.com} 
{\sc and} {\sc Nigel H. Scott}
\thanks{Email: n.scott@uea.ac.uk}\\[2pt]
{School of Mathematics, University of East Anglia,\\[2pt] Norwich Research Park,
 Norwich NR4 7TJ, UK}\\[6pt]
{\rm [Received on 21 November 2012;  Published online on 12 February 2013]}}
 
\maketitle
\pagestyle{headings}
\markboth{{\rm S. R. RICKABY AND N. H. SCOTT}}{{\rm SOFTENING  IN BIOLOGICAL TISSUE}}

\begin{abstract}
{In this paper we derive a model to describe the important inelastic features associated with the cyclic softening, often referred to as stress-softening, of soft biological tissue. The model developed here includes the notion of multiple stress-strain cycles with increasing values of the maximum strain. The model draws upon the similarities between the cyclic softening associated with  carbon-filled rubber vulcanizates and soft biological tissue. We give non-linear transversely isotropic models for the elastic response, stress relaxation, residual strain and creep of residual strain. These ideas are then combined with a transversely isotropic version of the Arruda-Boyce eight-chain model to develop a constitutive relation that is capable of accurately representing the multicyclic softening of soft biological tissue. To establish the validity of the model we have compared it with experimental data from three cyclic uniaxial test samples, one taken from the \textit{Manduca sexta} (tobacco hornworm) caterpillar and the other two samples taken from the human aorta, one in the  longitudinal and the other in the circumferential direction. The model was found to fit these experimental data extremely well.}
{Mullins effect, stress relaxation, creep of residual strain, biological tissue, transverse isotropy.
}
\end{abstract}

\section{Introduction} 

When a specimen of soft biological tissue is loaded, unloaded and then reloaded, the subsequent load required to produce the same deformation is smaller than that required during primary loading. This phenomenon is known as softening, often referred to as stress-softening, and can be described as a decay of elastic stiffness. Softening in soft biological tissue is particularly evident in muscle, skin and arteries.

\begin{figure}[ht]
\centering
\begin{tikzpicture}[scale=1.25, decoration={
markings,
mark=at position 6.5cm with {\arrow[black]{triangle 45};},
mark=at position 17.5cm with {\arrowreversed[black]{triangle 45};},
mark=at position 43cm with {\arrow[black]{triangle 45};},}
]
\draw[->] (0,0) -- (10,0)node[sloped,below,midway] {Stretch, $\lambda$} ;
\draw[->] (0,0) -- (0,5.5) node[sloped,above,midway] {Stress, $T_{11}$};
\draw [black](0,0) to[out=50,in=200] node [sloped,above] {} (6,3.5);
\draw [black](6,3.5) to[out=20,in=220] node [sloped,above] {} (9,5);
\draw [black] (2,0) to [out=6,in=260] node [sloped,above] {} (9,5);
\draw [black!40] 
(0.45,0) to[out=40,in=240] node [sloped,above] {} (9,4.5)
(2.45,0) to[out=6,in=256.5] node [sloped,below] {} (9,4.5);
\draw [dotted]
(0.9,0) to[out=40,in=238] node [sloped,below] {} (9,4.1)
(2.87,0) to[out=6,in=253] node [sloped,above] {} (9,4.1);
\draw [dashed] 
(1.22,0) to[out=40,in=236] node [sloped,below] {} (9,3.85)
(3.2,0) to[out=6,in=250] node [sloped,above] {} (9,3.85);
\draw [postaction={decorate}][loosely dotted,line width=0.01pt]
(0,0) to[out=50,in=200] node [sloped,above] {} (6,3.5)
(6,3.5) to[out=20,in=220] node [sloped,above] {} (9,5)
	(2,0) to[out=6,in=260] node [sloped,above] {} (9,5)
	(0.45,0) to[out=40,in=240] node [sloped,above] {} (9,4.5)
	(0.45,0) to[out=40,in=240] node [sloped,above] {} (9,4.5);
\coordinate [label=right:{$P_1^{\phantom{*}}, \,t_1^{\phantom{*}}$}] (B) at (9.0	,5);
\coordinate [label=right:{$P_2^{\phantom{*}}, \,t_2^{\phantom{*}}$}] (B) at (9.0	,4.5);
\coordinate [label=right:{$P_3^{\phantom{*}}, \,t_3^{\phantom{*}}$}] (B) at (9.0	,4.1);
\coordinate [label=right:{$P_4^{\phantom{*}}, \,t_4^{\phantom{*}}$}] (B) at (9.0	,3.75);
\coordinate [label=below:{$t_1^*$}] (B) at (2,-0.1);
\coordinate [label=below:{$t_2^*$}] (B) at (2.45,-0.1);
\coordinate [label=below:{$t_3^*$}] (B) at (2.85,-0.1);
\coordinate [label=below:{$t_4^*$}] (B) at (3.2,-0.1);
\coordinate [label=below:{$t_1^{**}$}] (B) at (0.45,-0.1);
\coordinate [label=below:{$t_2^{**}$}] (B) at (0.88,-0.1);
\coordinate [label=below:{$t_3^{**}$}] (B) at (1.26,-0.1);
\coordinate [label=below:{$P_1^*$}] (B) at (2,-0.5);
\coordinate [label=below:{$P_2^*$}] (B) at (2.45,-0.5);
\coordinate [label=below:{$P_3^*$}] (B) at (2.85,-0.5);
\coordinate [label=below:{$P_4^*$}] (B) at (3.2,-0.5);
\coordinate [label=below:{$P_1^{**}$}] (B) at (0.43,-0.5);
\coordinate [label=below:{$P_2^{**}$}] (B) at (0.88,-0.5);
\coordinate [label=below:{$P_3^{**}$}] (B) at (1.3,-0.5);
\coordinate [label=right:{$A$}] (B) at (4.2,3.3);
\coordinate [label=below:{$B$}] (B) at (5.3,1.4);
\coordinate [label=below:{$C$}] (B) at (4.75,2.37);
\coordinate [label=below:{$ $}] (B) at (0.9,0);
\coordinate [label=below:{$ $}] (B) at (1.2,0);
\coordinate [label=below:{$ $}] (B) at (1.5,0);
\coordinate [label=below:{$P_0^{\phantom{*}}$}] (O) at (0,0);
\draw plot[only marks, mark=square*, mark size=1.5pt, mark options={fill=black!30,draw=black!30}]  coordinates {(0.45,0)} ;
\draw plot[only marks, mark=diamond*, mark size=2pt, mark options={fill=black!30,draw=black!30}] coordinates {(2.0,0)} ;
\draw plot[only marks, mark=diamond*, mark size=2pt] coordinates {(2.45,0) (2.87,0) (3.2,0)} ;
\draw plot[only marks, mark=square*, mark size=1.5pt] coordinates {(0.9,0) (1.22,0) } ;
\end{tikzpicture}
\caption{Cyclic softening with residual strain.}
\label{fig:1}
\end{figure}
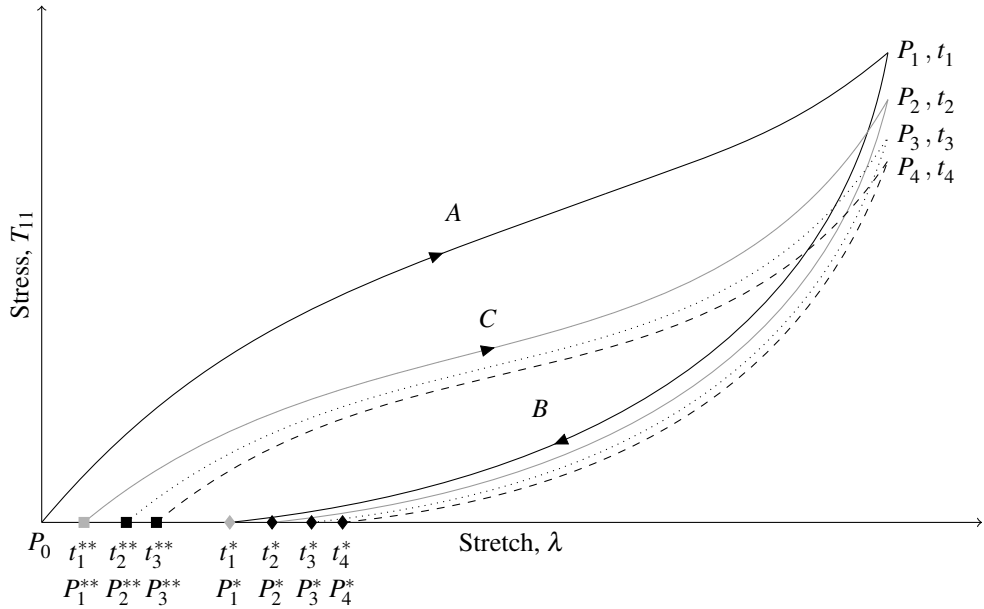
Figure \ref{fig:1} represents typical softening behaviour of a biological specimen under tension. The process starts from an unstressed virgin state at $P^{\phantom*}_0$ and the stress-strain relation follows path $A$, the primary loading path, until point $P^{\phantom*}_1$ is reached at a time $t^{\phantom*}_1$. At this point $P^{\phantom*}_1$, unloading of the biological specimen begins immediately and the stress-strain relation of the specimen follows the new path $B$ returning to the unstressed state at point $P^*_1$ and time $t^*_1$.  As a result of \textit{residual strain}, point  $P^*_1$ may not coincide with the origin $P^{\phantom*}_0$, but rather be at a position to the right of $P^{\phantom*}_0$,   marked by the grey diamond in Figure~\ref{fig:1}.  If reloading does not commence immediately but the material is kept in a state of zero stress then the amount of stretch reduces spontaneously, a phenomenon known as \textit{recovery} or \textit{creep of residual strain}.  This is allowed to continue until a point $P_1^{**}$, marked by a grey square, is reached, at a time $t_1^{**}$.
Reloading now commences and the stress-strain behaviour  follows the grey path $C$ until the same maximum strain is reached at point $P^{\phantom*}_2$ and time $t^{\phantom*}_2$. The fact that path $C$ does not coincide with path $B$ but lies above it constitutes the phenomenon of \textit{hysteresis}. Unloading of the biological specimen from the point $P^{\phantom*}_2$ starts immediately and the stress-strain relation follows the grey path to the unstressed state at point $P^*_2$.  As a result of \textit{stress relaxation} path $P^{\phantom*}_2P^*_2$ is situated below path $P^{\phantom*}_1P^*_1$.  This pattern  then continues throughout the unloading and reloading process as shown in Figure \ref{fig:1}.

Softening in biological material is analogous to softening in filled rubber vulcanizates. In vulcanized rubber this softening phenomenon is known as the Mullins effect, named after Mullins \cite{mullins1947}. Diani et al.\@ \cite{dianib} have written a recent review of this effect, detailing specific features associated with softening and providing a description of models developed to represent this effect. The comparison between the softening associated with soft biological tissue and filled vulcanized rubber has been discussed in detail by  Dorfmann et al.\@ \cite[Section 2]{dorfmann2007}.

It is observed experimentally that soft biological material is fibre reinforced.   This reinforcement gives the biological material an inherent anisotropic state, see Humphrey \cite[pages 264--267]{humphrey}. For example, tendons are considered to be transversely isotropic, see Humphrey  \cite[page 8]{humphrey}, and skin is considered to be orthotropic, see Lanir \& Fung \cite{lanir}.  Arteries exhibit cylindrical orthotropy with distinct circumferential and axial directions, see Humphrey  \cite[pages 264--267]{humphrey}.

Many authors have modelled cyclic stress softening  but most model a simplified version of this phenomenon in which one or more of the following inelastic features are excluded:

\begin{compactitem}
\item Hysteresis
\item Stress relaxation
\item Residual strain
\item Creep of residual strain
\end{compactitem}
In the numerical fit  to experimental data presented by Calvo et al.\@ \cite[Figure 9]{Calvo}, Pe\~na \& Doblar\'e \cite[Figures 1 and 2]{pena} all four inelastic features are neglected. The model comparison provided by Maher et al.\@ \cite[Figure 2]{Maher} includes only residual strain. Ehret et al.\@ \cite{ehret} model a preconditioned material whereby hysteresis, cyclic stress relaxation and creep of residual strain are excluded by previous working of the material. The preconditioned material presented by Dorfmann et al.\@ \cite{dorfmann2008} retains hysteresis, but excludes cyclic stress relaxation, residual strain and creep of residual strain.  Balzani et al.\@ \cite[Figure 3]{balzani} have developed a model for cyclic softening which replicates the broad stress softening features when compared with experimental data, though residual strain and creep of residual strain are excluded. 

Recently, Rickaby \& Scott \cite{rickaby} proposed a cyclic stress softening model for isotropic materials which models the inelastic features of hysteresis, stress relaxation, residual strain and creep of residual strain and achieved a high level of agreement with experiment.  Subsequently,  Rickaby \& Scott  \cite{rickaby1}  extended their isotropic model to the case of transverse isotropy.   This model was applied to carbon-filled rubber vulcanizates and good agreement with experiment was obtained.

The model developed in this paper is based on those of  Rickaby \& Scott \cite{rickaby1,rickaby} and accurately describes the cyclic stress softening of soft biological tissue for non-preconditioned experimental data over multiple stress-strain cycles with increasing values of maximum strain.

In Section \ref{sec:multi} we discuss the notion of stress softening  through multiple stress-strain cycles where the maximum strain is increased in successive cycles.   In Section \ref{sec:transversely}  we describe the transversely isotropic elastic response and in Section \ref{sec:softening} we discuss stress softening along the primary loading path and on the subsequent unloading and reloading paths. In Sections \ref{sec:relaxation} and \ref{sec:residual} we extend the  stress relaxation and creep of residual strain models of  Rickaby \& Scott  \cite{rickaby1, rickaby}  in order to incorporate multiple stress-strain cycles.  These individual components of our theory are then combined in Section \ref{sec:constitutive} to produce a general transversely isotropic model for stress softening, hysteresis, stress relaxation, residual strain and creep of residual strain.  This model is valid for multiple stress-strain cycles.   In Section \ref{sec:comparison} we compare the transversely isotropic model with experimental data in the case of uniaxial tension.  The experimental data is gained from three cyclic uniaxial test samples, one taken from the \textit{Manduca sexta} (tobacco hornworm) caterpillar and the other two samples taken from the human aorta in the longitudinal and circumferential directions. Finally, a discussion of the results is provided in Section \ref{sec:conclusion}.

\section{Multiple stress-strain cycles} 
\label{sec:multi}

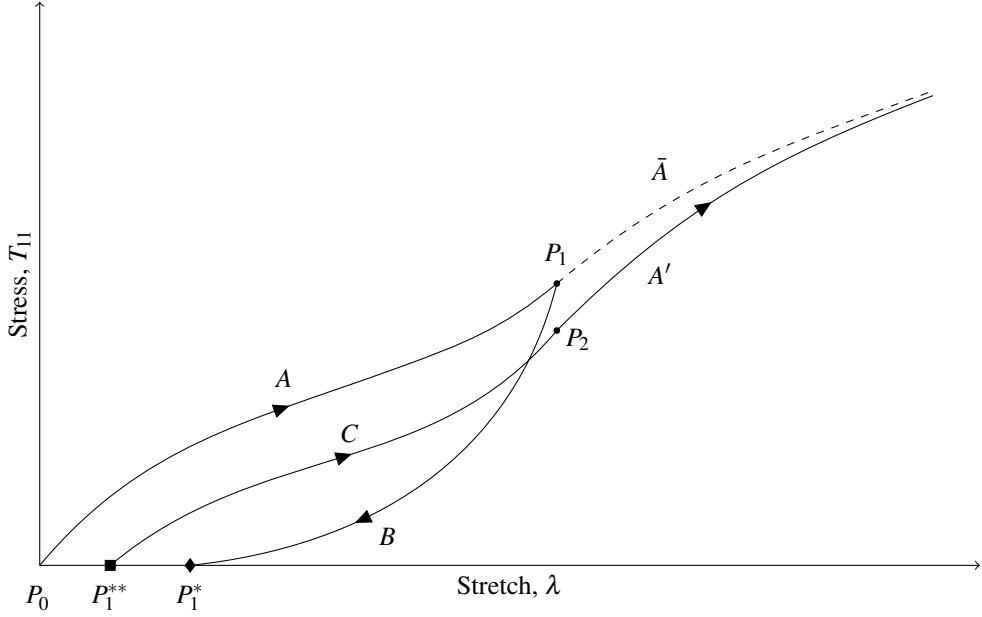
\begin{figure}[ht]
\centering
\begin{tikzpicture}[scale=1.25, decoration={
markings,
mark=at position 4cm with {\arrow[black]{triangle 45};},
mark=at position 11.5cm with {\arrow[black]{triangle 45};},
mark=at position 17cm with {\arrowreversed[black]{triangle 45};},
mark=at position 24.0cm with {\arrow[black]{triangle 45};}}
]
\draw[->] (-0.5,0) -- (9.5,0)node[sloped,below,midway] {Stretch, $\lambda$} ;
\draw[->] (-0.5,0) -- (-0.5,6) node[sloped,above,midway] {Stress, $T_{11}$};
\fill (5,3) circle (1pt);
\fill (5,2.5) circle (1pt);
\draw [postaction={decorate}]
(-0.5,0) to[out=50,in=200] node [sloped,above] {} (3.3,2.1)
(3.3,2.1) to[out=20,in=220] node [sloped,above] {} (5,3)
(0.25,0) to[out=40,in=230] node [sloped,above] {} (5,2.5)
(1.1,0) to[out=6,in=256] node [sloped,below] {} (5,3)
(5,2.5) to[out=45,in=201] node [sloped,below] {} (9,5);
\draw[dashed] (5,3) to[out=40,in=200] node [sloped,above] {} (9,5.05);
\coordinate [label=above:{${P}_1$}] (A) at (5,3.1);
\coordinate [label=right:{${P}_2$}] (A) at (5.0,2.4);
\coordinate [label=below:{$P_0^{\phantom{*}}$}] (O) at (-0.5,-0.1);
\coordinate [label=below:{${P}_1^*$}] (O) at (1.1,-0.1);
\coordinate [label=below:{${P}_1^{**}$}] (O) at (0.25,-0.1);
\coordinate [label=above:{$A$}] (A) at (2.1,1.8);
\coordinate [label=below:{$B$}] (O) at (3.2,0.5);
\coordinate [label=below:{$C$}] (O) at (2.8,1.6);
\coordinate [label=above:{$A'$}] (A) at (6.1,2.9);
\coordinate [label=above:{$\bar{A}$}] (A) at (6.1,4.0);
\draw plot[only marks, mark=square*, mark size=1.5pt] coordinates {(0.25,0)} ;
\draw plot[only marks, mark=diamond*, mark size=2pt] coordinates {(1.1,0)} ;
\end{tikzpicture}
\caption{Primary loading beyond the initial primary loading path.}
\label{fig:2}
\end{figure}

The paths $A$, $B$ and $C$ and the points $P_1$ and $P_2$ are exactly the same in each of Figures \ref{fig:1} and \ref{fig:2}.  However, in Figure \ref{fig:1} unloading commences after point  $P_2$ whereas in Figure \ref{fig:2} loading continues, so that the stretch and the stress both increase and the new path $A'$ is followed.   This new path is a continuation of path $C$ and is a second primary loading path.  The path $\bar{A}$, the dashed curve in Figure \ref{fig:2}, is the primary loading path that would have resulted if primary loading had continued beyond point $P_1$ (rather than commencing unloading and so following path $B$).  Path $\bar{A}$ is therefore a continuation of path $A$.   We argue that it is not possible for any part of path $A'$ to lie above path $\bar{A}$:  because of the softening on paths $B$ and $C$, for any given stretch the stress on $\bar{A}$ must be greater than that on $A'$.

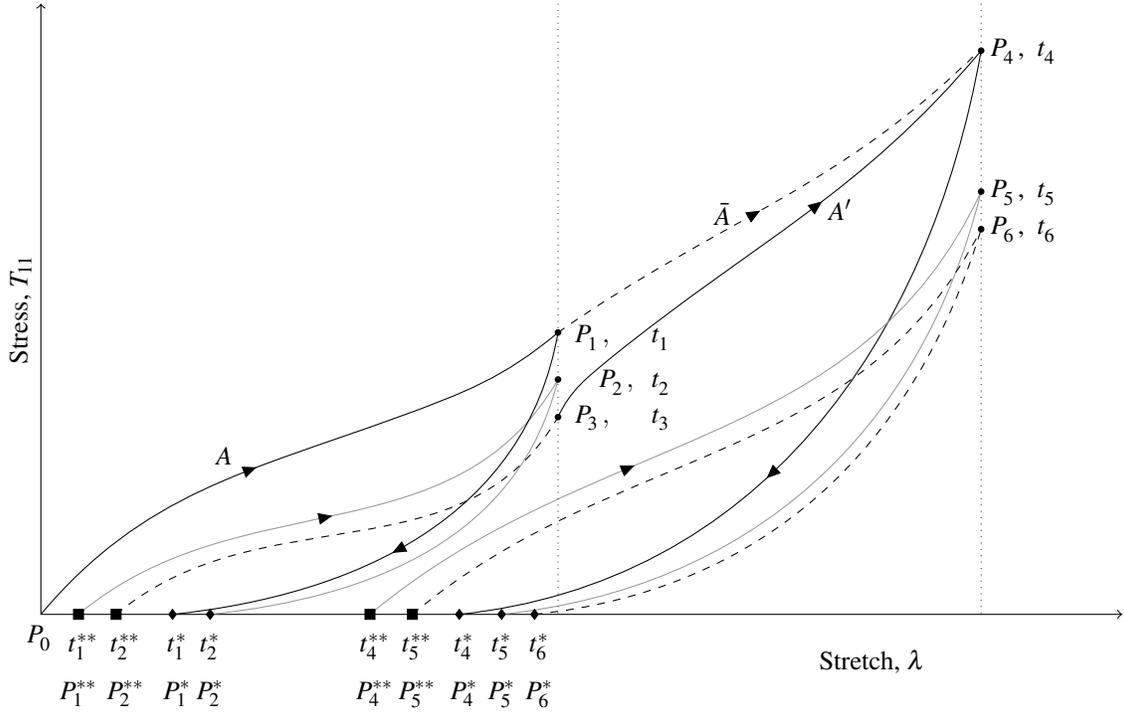
\begin{figure}[ht]
\centering
\begin{tikzpicture}[scale=1.25, decoration={
markings,
mark=at position 3.5cm with {\arrow[black]{triangle 45};},
mark=at position 11cm with {\arrowreversed[black]{triangle 45};},
mark=at position 21.5cm with {\arrow[black]{triangle 45};},}
]
\coordinate [label=left:{Stretch, $\lambda$}] (B) at (9,-0.5);
\draw[->] (-0.5,0) -- (11,0)node[sloped,below,midway] {} ;
\draw[->] (-0.5,0) -- (-0.5,6.5) node[sloped,above,midway] {Stress, $T_{11}$};
\draw [black!40]
(1.3,0) to[out=6,in=256] node [sloped,below] {} (5,2.5)
(-0.1,0) to[out=40,in=240] node [sloped,above] {} (5,2.5);
\draw [postaction={decorate}][loosely dotted,line width=0.01pt]
(-0.5,0) to[out=50,in=200] node [sloped,above] {} (3.3,2.1)
(3.3,2.1) to[out=20,in=220] node [sloped,above] {} (5,3)
(0.9,0) to[out=6,in=260] node [sloped,above] {} (5,3)
(0,0) to[out=6,in=260] node [sloped,above] {} (0,2.2)
(-0.1,0) to[out=40,in=240] node [sloped,above] {} (5,2.5);
\draw [black]
(-0.5,0) to[out=50,in=200] node [sloped,above] {} (3.3,2.1)
(3.3,2.1) to[out=20,in=220] node [sloped,above] {} (5,3)
(0.9,0) to[out=6,in=260] node [sloped,above] {} (5,3);
\draw [dashed]
(0.3,0) to[out=40,in=240] node [sloped,below] {} (5,2.1);
\fill (9.5,6) circle (1.pt);
\fill (5,3) circle (1.pt);
\fill (5,2.5) circle (1.pt);
\fill (5,2.1) circle (1.pt);
\draw [black]
(5.0,2.1) to[out=65,in=220] node [sloped,above] {} (5.55,2.72)
(5.55,2.72) to[out=39,in=230] node [sloped,above] {} (9.5,6)
(3.95,0) to[out=6,in=260] node [sloped,above] {} (9.5,6);
\draw [black!40]
(4.40,0) to[out=6,in=256] node [sloped,below] {} (9.5,4.5)
(3.00,0) to[out=40,in=245] node [sloped,above] {} (9.5,4.5);
\draw [postaction={decorate}][loosely dotted,line width=0.01pt]
(5.55,2.72) to[out=39,in=230] node [sloped,above] {} (9.5,6)
(3.95,0) to[out=6,in=260] node [sloped,above] {} (9.5,6)
(3.00,0) to[out=40,in=245] node [sloped,above] {} (9.5,4.5);
\draw [dashed]
(4.75,0) to[out=6,in=253] node [sloped,above] {} (9.5,4.1)
(3.45,0) to[out=40,in=243] node [sloped,below] {} (9.5,4.1);
\draw[dashed] (5,3) to[out=35,in=225] node [sloped,above] {} (9.5,6.02);
\draw [postaction={decorate}][loosely dotted,line width=0.01pt]
(0,0)--(0,0.3)
(5,3) to[out=35,in=225] node [sloped,above] {} (9.5,6.02);
\fill (9.5,4.5) circle (1.0pt);
\fill (9.5,4.1) circle (1.0pt);
\draw[black!80, dotted]
(9.5,6.5)--(9.5,0)
(5.0,6.5)--(5.0,0);
\coordinate [label=right:{$t_1^{\phantom{*}}$}] (B) at (5.9	,2.95);
\coordinate [label=right:{$t_2^{\phantom{*}}$}] (B) at (5.9	,2.48);
\coordinate [label=right:{$t_3^{\phantom{*}}$}] (B) at (5.9	,2.1);
\coordinate [label=right:{$P_1^{\phantom{*}},$}] (B) at (5.08	,2.95);
\coordinate [label=right:{$P_2^{\phantom{*}},$}] (B) at (5.34	,2.48);
\coordinate [label=right:{$P_3^{\phantom{*}},$}] (B) at (5.08	,2.1);
\coordinate [label=right:{$t_4^{\phantom{*}}$}] (B) at (10	,6);
\coordinate [label=right:{$t_5^{\phantom{*}}$}] (B) at (10	,4.5);
\coordinate [label=right:{$t_6^{\phantom{*}}$}] (B) at (10	,4.1);
\coordinate [label=right:{$P_4^{\phantom{*}}$,}] (B) at (9.5	,6);
\coordinate [label=right:{$P_5^{\phantom{*}}$,}] (B) at (9.5	,4.5);
\coordinate [label=right:{$P_6^{\phantom{*}}$,}] (B) at (9.5	,4.1);
\coordinate [label=below:{$t_1^*$}] (B) at (0.95,-0.1);
\coordinate [label=below:{$t_2^*$}] (B) at (1.3,-0.1);
\coordinate [label=below:{$P_2^*$}] (B) at (1.3,-0.6);
\coordinate [label=below:{$P_1^*$}] (B) at (0.95,-0.6);
\coordinate [label=below:{$t_4^*$}] (B) at (4,-0.1);
\coordinate [label=below:{$t_5^*$}] (B) at (4.4,-0.1);
\coordinate [label=below:{$t_6^*$}] (B) at (4.8,-0.1);
\coordinate [label=below:{$P_4^*$}] (B) at (4,-0.6);
\coordinate [label=below:{$P_5^*$}] (B) at (4.4,-0.6);
\coordinate [label=below:{$P_6^*$}] (B) at (4.8,-0.6);
\coordinate [label=below:{$t_1^{**}$}] (B) at (-0.05,-0.1);
\coordinate [label=below:{$t_2^{**}$}] (B) at (0.4,-0.1);
\coordinate [label=below:{$P_1^{**}$}] (B) at (-0.1,-0.6);
\coordinate [label=below:{$P_2^{**}$}] (B) at (0.4,-0.6);
\coordinate [label=below:{$t_4^{**}$}] (B) at (3.05,-0.1);
\coordinate [label=below:{$t_5^{**}$}] (B) at (3.5,-0.1);
\coordinate [label=below:{$P_4^{**}$}] (B) at (3.05,-0.6);
\coordinate [label=below:{$P_5^{**}$}] (B) at (3.5,-0.6);
\coordinate [label=above:{$A$}] (B) at (1.45,1.5);
\coordinate [label=above:{$A'$}] (B) at (8,4.1);
\coordinate [label=above:{$\bar{A}$}] (B) at (6.75,4.05);
\coordinate [label=below:{$P_0^{\phantom{*}}$}] (O) at (-0.5,0);
\draw plot[only marks, mark=diamond*, mark size=1.5pt] coordinates {(0.9,0) (1.3,0)} ;
\draw plot[only marks, mark=square*, mark size=1.5pt] coordinates {(-0.1,0) (0.3,0)} ;
\draw plot[only marks, mark=square*, mark size=1.5pt] coordinates {(3.00,0) (3.45,0)} ;
\draw plot[only marks, mark=diamond*, mark size=1.5pt] coordinates {(3.95,0) (4.40,0) (4.75,0)} ;
\end{tikzpicture}
\caption{Cyclic softening over two stress-strain cycles.}
\label{fig:3}
\end{figure}

These conclusions are borne out by the experimental data of  Diani et al.\@ \cite[Figure 1]{dianib}, which is very similar to Figures  \ref{fig:2} and   \ref{fig:3} here.

The stress-strain behaviour of a biological specimen undergoing multiple stress-strain cycles is illustrated in Figure \ref{fig:3}.  Two cycles are shown.  In the first cycle the specimen is loaded along path $A$ to the maximum stretch value 
$\lambda_{\C\_1}$, denoted by $\lambda_\M$ in  Rickaby \& Scott \cite{rickaby1, rickaby}. It is then unloaded and reloaded repeatedly, each time to the maximum stretch value $\lambda_{\C\_1}$, achieved at the points $P_1$, $P_2$ and $P_3$ in turn.  This completes cycle one.  Cycle two now commences at point $P_3$ when the stress is increased further rather than being allowed to relax back to zero.  The stretch also increases, to a new maximum value 
$\lambda_{\C\_2}$, and then a further cycle of unloading and reloading is carried out, each time to the stretch $\lambda_{\C\_2}$, achieved at the points $P_4$, $P_5$ and $P_6$ in turn. This completes cycle two.

\section{Transversely isotropic elastic response} 
\label{sec:transversely}
For a transversely isotropic, fibre-reinforced material the strain energy function takes the form
\begin{equation}
W=W(I_1,I_2,I_3,I_4,I_5),
\label{eq:3.1w}
\end{equation}
where the material invariants are defined by
\begin{align*}
I_1&=\tr\textbf{C}, \quad
I_2=\tfrac{1}{2}\big\{(\tr\textbf{C})^2-\tr (\textbf{C}^2)\big\}, \quad
I_3=\det\textbf{C},\\
I_4&=\textbf{A}\cdot(\textbf{C}\textbf{A}),\quad
I_5=\textbf{A}\cdot(\textbf{C}^2\textbf{A}),
\end{align*}
in which $\textbf{C}=\textbf{F}^\T\textbf{F}$ is the right Cauchy-Green strain tensor,  $\textbf{F}$ is the deformation gradient and $\textbf{A}$ is a unit vector in the direction of transverse isotropy in the reference configuration.  

It is common to regard biological tissue as being incompressible,   so that $I_3=1$ for all deformations and so $I_3$ may be omitted from (\ref{eq:3.1w}).   We omit also $I_2$  from (\ref{eq:3.1w}) because the models of elasticity which we employ, namely Arruda \& Boyce  \cite{arruda}  and Kuhl et al.\@ \cite{kuhl}, do not involve it.  Merodio \& Ogden 
 \cite{merodio} regard the invariant $I_5$ as being related to shearing orthogonal to the fibre direction.  If we assume that  any such shearing is minimal then we may further omit  $I_5$ from (\ref{eq:3.1w}). 
 The strain energy (\ref{eq:3.1w}) then reduces to
\begin{equation}
W=W(I_1, I_4)
\label{eq:3.2w}
\end{equation}
leading to the Cauchy stress 
\begin{align}
\textbf{T}^{\mathscr{E_{\ti}}}(\lambda) =&\, -p\textbf{I}+2\bigg\{\frac{\partial{W}}{\partial I_1}
\textbf{B}+ \frac{\partial{W}}{\partial
I_4}\bm{a}\otimes\bm{a}\bigg\},
\label{eq:3.3w}
\end{align}
where $p$ is an arbitrary pressure resulting from incompressibility, $\textbf{I}$ is the identity matrix, $\textbf{B}=\textbf{F}\textbf{F}^\T$ is the left Cauchy-Green strain tensor, $\otimes$ denotes a dyadic product and $\bm{a}=\textbf{F}\textbf{A}$.

As a model for the elastic behaviour of biological tissue we employ the transversely isotropic eight-chain model of  Rickaby \& Scott  \cite{rickaby1} and Kuhl et al.\@  \cite{kuhl}, which is itself based on the original eight-chain model of Arruda \& Boyce  \cite{arruda} for isotropic elasticity.  To this end we define the quantities
\[
 \gamma=\sqrt{\frac{I_4 +\left(I_1-I_4\right)\alpha^2}{N(1+2\alpha^2)}}, \quad \beta = \mathscr{L}^{-1}(\gamma),
\]
in which ${N}$ is the number of links forming a single
polymer chain, $\mathscr{L}^{-1}(\cdot)$ is the inverse of the Langevin function $\mathscr{L}(x) = \coth x - 1/x$ and
$\alpha>0$ is the aspect ratio parameter, see Kuhl et al.\@  \cite{kuhl}, with $\alpha=1$ corresponding to an isotropic material. 

The appropriate form of the strain energy function that we adopt here is a combination of the transversely isotropic Arruda-Boyce eight chain model coupled with a term linear in $I_4$:
\begin{align}
W_{\textrm{A-B}} &= \mu N \left\{\gamma \beta+\log \left(\frac{\beta}{\sinh \beta}\right) \right\}-\frac{1}{2}h_{4}\left(I_4-1\right),
\label{eq:3.4w}
\end{align}
where $\mu$ is the ground state shear modulus 
and 
\[  h_4 =\mu \frac{1-\alpha^2}{1+2\alpha^2} \sqrt N  \mathscr{L}^{-1}\left(\sqrt{\frac{1} N }\right) \] 
is a constant chosen so that the  stress vanishes in the  undeformed state. 

A fully transversely isotropic contribution to the strain energy is furnished by %
\begin{equation}
W_{\ti}=\tfrac{1}{2} s_1 I_4^{-1}(I_4-1)^2+\tfrac{1}{2}s_2 I_4^{-1} (I^{\frac{1}{2}}_4-1)^2,
\label{eq:3.5w}
\end{equation}
where $s_1$ and $s_2$ are constants.  This strain energy gives rise to a zero stress in the reference configuration.

 From equations (\ref{eq:3.3w}), (\ref{eq:3.4w}) and (\ref{eq:3.5w}) we obtain the transversely isotropic elastic stress 
\begin{align}
\textbf{T}^{\mathscr{E}_{\ti}}(\lambda) =&\, -p\mathbf{I} + 
\mu\frac{\alpha^2}{1+2\alpha^2} \gamma^{-1}\beta\, \mathbf{B} + 
\left(\mu\frac{1 - \alpha^2}{1+2\alpha^2} \gamma^{-1}\beta - h_{4}\right) \bm{a}\otimes\bm{a} \nonumber \\
 & + I^{-2}_4  \bigg({s_1}(I^{2}_4 -1)+{s_2}(I^{\frac{1}{2}}_4- 1)\bigg)  \bm{a}\otimes\bm{a}.
\label{eq:3.6w}
\end{align}

We have chosen to use the Arruda-Boyce model though we could equally well have used the simpler Gent \cite{gent} model. Boyce \cite{boyce1996} and Horgan \& Saccomandi  \cite{horgan2002} discuss the  similarity between these two models.  Han et al.\@ \cite{han} have criticized the Arruda-Boyce model, though it has been shown to give a very good fit to experimental data, see Boyce \& Arruda \cite{boyce}.

\section{Softening functions} 
\label{sec:softening}

\subsection{Softening on the primary loading path}
\label{sec:initial}

For  biological tissue it is observed experimentally, see Holzapfel et al.\@ \cite{holzapfel2000},  that during initial primary loading pronounced softening occurs but only at small deformations.   To describe this feature we adopt here the  softening function
\[
\zeta_0(\lambda)=1-\frac{1}{r_0}\left\{\tanh\left(\frac{\lambda_{\C\_1}-{\lambda}}{ b_0}\right)\right\}^{{1}/{\vartheta_0}}, 
\]   
 for $ 1<\lambda\leq\lambda_{\C\_1}$, originally proposed for carbon-filled rubber vulcanizates by  Rickaby \& Scott  \cite{rickaby1} who wrote $\lambda_\M$ in place of  $\lambda_{\C\_1}$.   Here, $r_0$, $b_0$ and  $\vartheta_0$ are positive constants and    $\lambda_{\C\_1}$ is the greatest stretch achieved.
 For  primary loading this softening function may be coupled with the isotropic  elastic stress  (\ref{eq:3.6w}) to give
\begin{align}
\textbf{T}^{\mathscr{E}_{\ti}}(\lambda)=\zeta_{0}(\lambda)\Bigg[&\, -p\mathbf{I} + 
\mu\frac{\alpha^2}{1+2\alpha^2} \gamma^{-1}\beta\, \mathbf{B} + 
\left(\mu\frac{1 - \alpha^2}{1+2\alpha^2} \gamma^{-1}\beta - h_{4}\right) \bm{a}\otimes\bm{a} \nonumber \\
 & \quad + I^{-2}_4  \bigg({s_1}(I^{2}_4 -1)+{s_2}(I^{\frac{1}{2}}_4- 1)\bigg)  \bm{a}\otimes\bm{a}\Bigg].
\label{eq:4.1w}
\end{align}
Because softening on the primary loading path occurs only at  small strains we must have $\zeta_0(\lambda) \approx 1$  for larger strains.

\subsection{Softening on the unloading and reloading paths}

Motivated by Dorfmann \& Ogden \cite{dorfmann2003,dorfmann},  Rickaby \& Scott  \cite{rickaby1,rickaby} introduced the following softening function
in order to model stress softening on the unloading and reloading paths, i.e. paths $B$ and $C$ of  
Figure \ref{fig:1}:
\begin{equation}
\zeta_\omega(\lambda) = 1-\frac{1}{r_\omega}\left\{\tanh\left(\frac{W_{\C\_1}-{W}}{\mu b_\omega}\right)\right\}^{{1}/{\vartheta_\omega}}, 
\label{eq:4.2w}
\end{equation}
where $W_{\C\_1}$ is the maximum strain energy  achieved at the strain $\lambda_{\C\_1}$ on the primary loading path $A$ of Figure \ref{fig:1} and $W$ is the current value of the strain energy.    The quantities $r_\omega$,  $b_\omega$ and $\vartheta_\omega$ are positive dimensionless material constants labelled by
 \begin{equation}
\omega=\left\{ \begin{array}{clrr}
1 & \textrm{for unloading}, \\[2mm]
2 & \textrm{for reloading}. \\
\end{array}\right.
 \label{eq:4.3w}
 \end{equation}

The softening function  (\ref{eq:4.2w}) has the property that
 \begin{equation}
\textbf{T} =  \zeta_\omega(\lambda) \textbf{T}^{\mathscr{E}_{\ti}}(\lambda),
 \label{eq:4.4w}
 \end{equation}
providing a connection between the Cauchy stress $\textbf{T}$ in the    unloading and reloading phases of the material and the Cauchy stress  $\textbf{T}^{\mathscr{E}_{\ti}}(\lambda)$ in the primary loading phase of a transversely isotropic elastic parent material.

In straining the material beyond point $P_3$ in Figure \ref{fig:3} to a point $P_4$, a new maximum stretch value $\lambda_{\C\_2}>\lambda_{\C\_1}$ is imposed. The softening function is now  dependent on this new maximum stretch value. This pattern  then continues for cyclic loading to higher stretches which are denoted by $\lambda_{\C\_n} >\lambda_{\C\_(n-1)},\; n=2,3,\dots$.  Then $W_{\C\_1}$ in equation (\ref{eq:4.2w}) can be replaced by $W_{\C\_n}$ to give
\begin{equation}
\zeta_{n,\omega}(\lambda) = 1-\frac{1}{r_\omega}\left\{\tanh\left(\frac{{W_{\C\_n}}-{W}}{\mu b_\omega}\right)\right\}^{{1}/{\vartheta_\omega}},
 \label{eq:4.5w}
\end{equation}
where $n$ counts the cycles and $\omega$ is defined in equation (\ref{eq:4.3w}).

\section{Stress relaxation} 
\label{sec:relaxation}

Suppose a material body is deformed in some way by applied stresses and is then held in the same state of deformation over a period of time by applied stress. Stress relaxation is said to occur if the stress needed to maintain this fixed deformation decreases over the period of time.

Figure \ref{fig:1} represents a cyclically loaded and unloaded biological specimen with primary loading occurring along  path $P_0^{\phantom{*}}P_1^{\phantom{*}}$ up to the point $P_1^{\phantom{*}}$,  which is reached at  time $t_1$. We postulate that stress-relaxation commences from the point of initial loading at time $t_0$ and follows the primary loading path $P_0^{\phantom{*}}P_1^{\phantom{*}}$ up to the point $P_1^{\phantom{*}}$. Stress-relaxation then follows the unloading  path $P_1^{\phantom{*}}P_1^{*}$  back to the position~$P_1^{*}$ of zero stress, which is reached at time $t^*_1$. Stress-relaxation is assumed to continue in the stress free state.  On the  reloading  path $P_1^{**}P_2^{\phantom{*}}$ stress-relaxation proceeds until point $P_2$ is reached, at time $t_2$.    Stress-relaxation continues as we follow the grey  unloading path $P_2^{\phantom{*}}P_2^{*}$ to the position $P_2^{*}$  of zero stress, reached at time $t^*_2$. This pattern then continues throughout the unloading and reloading process.

 For repeated cyclic loading to the same stretch value $\lambda_{\C\_1}$, as shown in Figure  \ref{fig:1},  Rickaby \& Scott \cite{rickaby1} derived the following  transversely isotropic stress-relaxation function:
\begin{align}
\textbf{T}^{\mathscr{R}_{\ti}}(\lambda,t)=&\,-p\textbf{I} +\bigg[ {A}_0+\frac{1}{2} {A}_1(t)[I_1-3]-
{A}_2(t)\bigg]{\textbf{B}}+ {A}_2(t){\textbf{B}}^{2}\nonumber\\
&+{A}_4(t)(I_4-1)\bm{a}\otimes \bm{a}+{A}_5(t)(I_5-1)\left(\bm{a}\otimes\textbf{B}\bm{a}+ \textbf{B}\bm{a}\otimes \bm{a}\right),
\label{eq:5.1w}
\end{align}
for  $ t>t_0$, with $\textbf{T}^{\mathscr{R}_{\ti}}(\lambda,t)$ vanishing for $t\leq t_0$. In equation (\ref{eq:5.1w}), ${A}_0$ is a material constant and ${A}_l(t)$, where $l\in\left\{1,2,4,5\right\}$, are material functions defined by
\begin{equation}
A_l(t)=\left\{\begin{array}{llll}
\breve{A}_l(\phi_0(t-t_1))                                      & \textrm{primary loading}, & t_0^{\phantom{*}}\leq t\leq t_1^{\phantom{*}}, &  \textrm{path}\;\; P_0^{\phantom{*}}P_1^{\phantom{*}}\\[2mm]
\breve{A}_l(\phi_1(t-t_1))& \textrm{unloading}, & t_1^{\phantom{*}}\leq t\leq t_1^*, &  \textrm{path}\;\; P_1^{\phantom{*}}P_1^*\\[2mm]
\breve{A}_l(\phi_1(t-t_1))& \textrm{stress free}, & t^*_1\leq t\leq t_1^{**}, &  \textrm{path}\;\; P_1^*P_1^{**}\\[2mm] 
\breve{A}_l(\phi_2(t-t_1))& \textrm{reloading}, & t^{**}_1\leq t\leq t_2^{\phantom{*}}, &  \textrm{path}\;\; P_1^{**}P_2^{\phantom{*}}\\[2mm]
 \;\; \dots&\;\; \dots&\;\; \dots&\dots
\end{array}\right.
\label{eq:5.2w}
\end{equation}
Each function  $\breve{A}_l(t)$ vanishes for $t\leq t_0$ and is continuous. In equation (\ref{eq:5.2w}), $\phi_0$, $\phi_1$ and $\phi_2$ are continuous functions of time. For simplicity, on the stress-free paths we also employ $\phi_1$ as the argument for $A_l(t)$.

If the material is now strained beyond the value $\lambda_{\C\_1}$ of stretch,  path $C$  continues onto path $A'$ as shown in Figure \ref{fig:2}. In the present model we assume that stress relaxation, given by equation (\ref{eq:5.1w}), continues to evolve with time on the primary loading path $A'$, i.e. path $P_3P_4$, but at a different rate from that of unloading and reloading because the functions $A_l(t)$ are different along different paths.   In straining the material beyond point $P_3$ to a point $P_4$ as shown in Figure \ref{fig:3} a new maximum stretch value $\lambda_{\C\_2}$ is imposed. 

For the multiple stress-strain cycles shown in Figure \ref{fig:3}, the function $A_l(t)$, given by equation (\ref{eq:5.2w}),  becomes
\begin{equation}
A_l(t)=\left\{\begin{array}{llll}
\breve{A}_l(\phi_0(t-t_1))                                      & \textrm{primary loading}, & t_0^{\phantom{*}}\leq t\leq t_1^{\phantom{*}}, &  \textrm{path}\;\; P_0^{\phantom{*}}P_1^{\phantom{*}}\\[2mm]
\breve{A}_l(\phi_{ {1,1}}(t-t_1))& \textrm{unloading}, & t_1^{\phantom{*}}\leq t\leq t_1^*, &  \textrm{path}\;\; P_1^{\phantom{*}}P_1^*\\[2mm]
\breve{A}_l(\phi_{ {1,1}}(t-t_1))& \textrm{stress free}, & t^*_1\leq t\leq t_1^{**}, &  \textrm{path}\;\; P_1^*P_1^{**}\\[2mm] 
\breve{A}_l(\phi_{ {1,2}}(t-t_1))& \textrm{reloading}, & t^{**}_1\leq t\leq t_2^{\phantom{*}}, &  \textrm{path}\;\; P_1^{**}P_2^{\phantom{*}}\\[2mm]
 \;\; \dots&\;\; \dots&\;\; \dots&\;\; \dots \\[2mm]
\breve{A}_l(\phi_{ {2,3}}(t-t_1))& \textrm{primary loading}, & t_3^{\phantom{*}}\leq t\leq t_4^{\phantom{*}}, &  \textrm{path}\;\; P_3^{\phantom{*}}P_4^{\phantom{*}}\\[2mm]
 \;\; \dots&\;\; \dots&\;\; \dots&\;\; \dots 
\end{array}\right.
\label{eq:5.3w}
\end{equation}
in which $\phi_{n,\omega}$ are  continuous functions of time, with $n$ counting the number of cycles and $\omega$ being defined by equation (\ref{eq:4.3w}). Note the occurrence of the new functions $\phi_{n,3}$ which arise because of the new primary loading paths, for example, $\phi_{2,3}$ relates to path  $P_3P_4$ of  Figure \ref{fig:3}.  

In the literature on cyclic softening we have been unable to identify any other model that  takes  into consideration the effect of stress relaxation associated with  multicyclic stress-strain  loading.

\section{Creep of residual strain} 
\label{sec:residual}

For  cyclic loading to the same maximum stretch value, $\lambda_{\C\_1}$, Rickaby \& Scott  \cite{rickaby1} derived the following expression for the creep stress causing the   creep of residual strain in the transversely isotropic case:
\begin{equation}
\textbf{T}^{\mathscr{C}_{\ti}}(\lambda, t)=-p \textbf{I}+\left\{d_\omega\left[\sqrt{N}\gamma-1\right]^{-1}
\left\{1+\left[\tanhs a(t)\right]^{a_1}\right\}\right\} \mathbf{B}, 
\label{eq:6.1w}
\end{equation}
for $t > t_1$ and $\lambda>1$, with  $\textbf{T}^{\mathscr{C}_{\ti}}(\lambda,t)$ vanishing for $t\leq t_1$. In equation (\ref{eq:6.1w}), $a_1$ and $d_\omega$ are material constants,  $d_1$  for    unloading and $d_2$  for    reloading, with $d_2\leq d_1$. 

The function $a(t)$ is defined by
\begin{equation}
a(t)=\left\{\begin{array}{llll}
   0                                      & \textrm{primary loading}, & t_0^{\phantom{*}}\leq t\leq t_1^{\phantom{*}}, &  \textrm{path}\;\; P_0^{\phantom{*}}P_1^{\phantom{*}}\\[2mm]
\breve{a}(\Phi_1(t-t_1))& \textrm{unloading}, & t_1^{\phantom{*}}\leq t\leq t_1^*, &  \textrm{path}\;\; P_1^{\phantom{*}}P_1^*\\[2mm]
\breve{a}(\Phi_1(t-t_1))& \textrm{stress free}, & t^*_1\leq t\leq t_1^{**}, &  \textrm{path}\;\; P_1^*P_1^{**}\\[2mm] 
\breve{a}(\Phi_2(t-t_1))& \textrm{reloading}, & t^{**}_1\leq t\leq t_2^{\phantom{*}}, &  \textrm{path}\;\; P_1^{**}P_2^{\phantom{*}}\\[2mm]
 \;\; \dots&\;\; \dots&\;\; \dots&\;\; \dots
\end{array}\right.
\label{eq:6.2w}
\end{equation}
where $\Phi_1$ and $\Phi_2$ are continuous functions of time. For simplicity, on the stress-free paths we also employ $\Phi_1$ as the argument for $a(t)$.

If the material is now stretched beyond the stretch value $\lambda_{\C\_1}$, path $C$ is extended onto  path $A'$ as shown in Figure \ref{fig:2}.   We conjecture that  the residual strain function  continues to evolve with time, but at a rate different from that of    unloading and reloading. In straining the material beyond point $P_3$ to a point $P_4$ a new maximum stretch value $\lambda_{\C\_2} >\lambda_{\C\_1} $ is imposed. This pattern is then repeated for cyclic loading to higher stretches which are denoted by $\lambda_{\C\_n} >\lambda_{\C\_(n-1)},\; n=2,3,\dots$. The  creep stress (\ref{eq:6.1w})  now becomes dependent on these new maximum stretch values. This feature can be included within the model by regarding $d_\omega$ in equation (\ref{eq:6.1w}) as a function of the applied maximum stretch $\lambda_{\C\_n}$. Then for deformations involving multiple stress-strain cycles the creep stress of residual strain (\ref{eq:6.1w}) becomes
\begin{equation}
\textbf{T}^{\mathscr{C}_{\ti}}(\lambda, t)=-p \textbf{I}+\left\{d_\omega(\lambda_{\C\_n})\left[\sqrt{N}\gamma-1\right]^{-1}\left\{1+\left[\tanhs a(t)\right]^{a_1}\right\}\right\} \mathbf{B}, 
\label{eq:6.3w}
\end{equation}
for $t > t_1$ and $\lambda>1$. 

Now, in place of (\ref{eq:6.2w}),  $a(t)$ is defined by

\begin{equation}
a(t)=\left\{\begin{array}{llll}
   0                                      & \textrm{primary loading}, & t_0^{\phantom{*}}\leq t\leq t_1^{\phantom{*}}, &  \textrm{path}\;\; P_0^{\phantom{*}}P_1^{\phantom{*}}\\[2mm]
\breve{a}(\Phi_{ {1,1}}(t-t_1))& \textrm{unloading}, & t_1^{\phantom{*}}\leq t\leq t_1^*, &  \textrm{path}\;\; P_1^{\phantom{*}}P_1^*\\[2mm]
\breve{a}(\Phi_{ {1,1}}(t-t_1))& \textrm{stress free}, & t^*_1\leq t\leq t_1^{**}, &  \textrm{path}\;\; P_1^*P_1^{**}\\[2mm] 
\breve{a}(\Phi_{ {1,2}}(t-t_1))& \textrm{reloading}, & t^{**}_1\leq t\leq t_2^{\phantom{*}}, &  \textrm{path}\;\; P_1^{**}P_2^{\phantom{*}}\\[2mm]
 \;\; \dots&\;\; \dots&\;\; \dots&\;\; \dots\\[2mm]
\breve{a}(\Phi_{ {2,3}}(t-t_1))& \textrm{primary loading}, & t_3^{\phantom{*}}\leq t\leq t_4^{\phantom{*}}, &  \textrm{path}\;\; P_3^{\phantom{*}}P_4^{\phantom{*}}\\[2mm]
 \;\; \dots&\;\; \dots&\;\; \dots&\;\; \dots
\end{array}\right.
\label{eq:6.4w}
\end{equation}
in which $\Phi_{n,\omega}$ are  continuous functions of time. The new functions $\Phi_{n,3}$ arise from the new primary loading paths. 

In the literature on cyclic softening we have been unable to identify any other model that  takes  into consideration the effect of creep of residual strain associated with  multicyclic stress-strain  loading. 

\section{Constitutive model} 
\label{sec:constitutive}
The total stress $\textbf{T}$ in the material is now obtained by combining together all the different stresses discussed in previous sections to obtain
\begin{equation}
\mathbf{T}=\left\{\begin{array}{llll}
 \,\,\,\zeta_0(\lambda)\textbf{T}^{\mathscr{E_{\ti}}+\mathscr{R_{\ti}}}(\lambda,t),& \textrm{primary loading}, & t_0^{\phantom{*}}\leq t\leq t_1^{\phantom{*}}, & \textrm{path}\;\; P_0^{\phantom{*}}P_1^{\phantom{*}}\\[2mm]                 
\zeta_{1,1}(\lambda)\textbf{T}^{\mathscr{E_{\ti}}+\mathscr{R_{\ti}}+\mathscr{C_{\ti}}}(\lambda, t),& \textrm{unloading}, & t_1^{\phantom{*}}\leq t\leq t_1^{*}, &  \textrm{path}\;\; P_1^{\phantom{*}}P_1^*\\[2mm]
\phantom{\zeta_1(\lambda)\big\{}{\bf 0} & \textrm{stress free}, & t^*_1\leq t\leq t_1^{**}, &  \textrm{path}\;\; P_1^*P_1^{**}\\[2mm] 
\zeta_{1,2}(\lambda)\textbf{T}^{\mathscr{E_{\ti}}+\mathscr{R_{\ti}}+\mathscr{C_{\ti}}}(\lambda, t),& \textrm{reloading}, & t^{**}_1\leq t\leq t_2^{\phantom{*}}, &  \textrm{path}\;\; P_1^{**}P_2^{\phantom{*}}\\[2mm]
 \;\; \dots&\;\; \dots&\;\; \dots&\;\; \dots\\[2mm]
\phantom{\zeta_{2,2}(\lambda)}\textbf{T}^{\mathscr{E_{\ti}}+\mathscr{R_{\ti}}+\mathscr{C_{\ti}}}(\lambda, t),& \textrm{primary loading}, & t_3^{\phantom{*}}\leq t\leq t_4^{\phantom{*}}, &  \textrm{path}\;\; P_3^{\phantom{*}}P_4^{\phantom{*}}\\[2mm]
 \;\; \dots&\;\; \dots&\;\; \dots&\;\; \dots
 \end{array}\right. 
 \label{eq:7.1w}
 \end{equation}
in which for notational convenience we have defined  stresses
\begin{align*}
  \textbf{T}^{\mathscr{E_{\ti}}+\mathscr{R_{\ti}}}(\lambda, t)  =&\, \textbf{T}^{\mathscr{E}_{\ti}}(\lambda) + \textbf{T}^{\mathscr{R}_{\ti}}(\lambda, t),\\[4pt]
  \textbf{T}^{\mathscr{E_{\ti}}+\mathscr{R_{\ti}}+\mathscr{C_{\ti}}}(\lambda, t)  =&\, \textbf{T}^{\mathscr{E}_{\ti}}(\lambda) + \textbf{T}^{\mathscr{R}_{\ti}}(\lambda, t) + \textbf{T}^{\mathscr{C}_{\ti}}(\lambda, t),
\end{align*} 
where $\textbf{T}^{\mathscr{E}_{\ti}}(\lambda)$, $\textbf{T}^{\mathscr{R}_{\ti}}(\lambda, t)$ and $ \textbf{T}^{\mathscr{C}_{\ti}}(\lambda, t)$ are given by equations  (\ref{eq:4.1w}), (\ref{eq:5.1w}) and (\ref{eq:6.3w}), respectively.  Note that the stress on the second primary loading path $P_3P_4$ is not multiplied by the primary softening function because 
$\zeta_0(\lambda) \approx 1$ for the larger values of strain encountered on this path, see Section \ref{sec:initial}.

To the best of our knowledge the effects of stress relaxation and creep of residual strain in relation to softening during multicyclic stress-strain  loading have not previously been considered in the literature and so  equation (\ref{eq:7.1w}) for the stress has not previously been exhibited.

The transversely isotropic constitutive model is then given by
\begin{align}
\textbf{T}^{\ti} =&\,\left[1-\frac{1}{r_\omega}\left\{\tanh\left(\frac{W_{\C\_1}-{W}}{\mu b_\omega}\right)\right\}^{{1}/{\vartheta_\omega}}\right]\times\nonumber\\
&\times\Bigg\{ -p\mathbf{I} + 
\mu\frac{\alpha^2}{1+2\alpha^2} \gamma^{-1}\beta\mathbf{B} + 
\left(\mu\frac{1 - \alpha^2}{1+2\alpha^2} \gamma^{-1}\beta - h_{4}\right) \bm{a}\otimes\bm{a}\nonumber\\
&\qquad+\bigg[ {A}_0+\frac{1}{2} {A}_1(t)(I_1-3) -
{A}_2(t)\bigg]{\textbf{B}}+ {A}_2(t){\textbf{B}}^{2}\nonumber\\
&\qquad+{A}_4(t)(I_4-1)\bm{a}\otimes \bm{a}+{A}_5(t)(I_5-1)\left(\bm{a}\otimes\textbf{B}\bm{a}+ \textbf{B}\bm{a}\otimes \bm{a}\right)
\nonumber\\
&\qquad+d_\omega(\lambda_{\C\_n})\left[\sqrt{N}\gamma-1\right]^{-1}\left\{1+\left[\tanhs a(t)\right]^{a_1}\right\} \mathbf{B}\nonumber\\
&\qquad+ I^{-2}_4 \left({s_1}(I^{2}_4 - 1)+{s_2}(I^{\frac{1}{2}}_4- 1)\right)\bm{a}\otimes\bm{a}\Bigg\}.
\label{eq:7.2w}
\end{align}

\section{Comparison with experimental data in uniaxial tension} 
\label{sec:comparison}

\subsection{Caterpillar muscle}

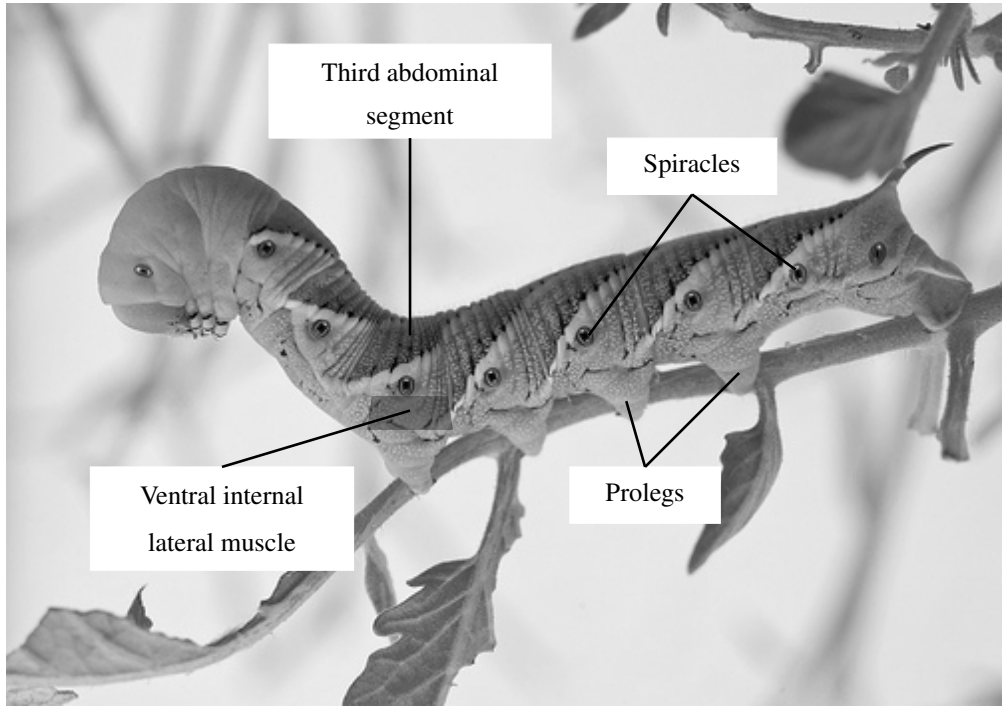
\begin{figure}[ht]
\centering
\begin{tikzpicture}[scale=1.25]
\node (0,0) {\includegraphics[angle=180, scale=2.5]{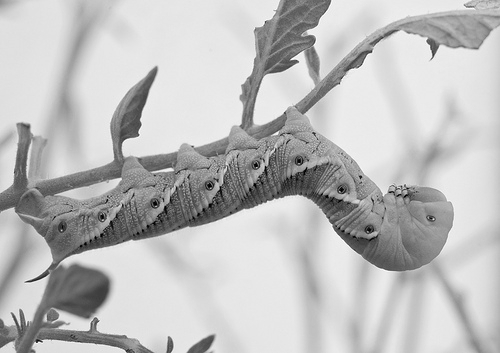}};
\filldraw[fill=white,draw=white](0.7,-1.8)--(0.7,-1.2)--(2.3,-1.2)--(2.3,-1.8)--(0.7,-1.8);
\filldraw[fill=white,draw=white](1.1,1.7)--(1.1,2.3)--(2.9,2.3)--(2.9,1.7)--(1.1,1.7);
\filldraw[fill=white,draw=white](-2.5,2.3)--(-2.5,3.3)--(0.5,3.3)--(0.5,2.3)--(-2.5,2.3);
\filldraw[fill=white,draw=white](-4.4,-2.3)--(-4.4,-1.2)--(-1.6,-1.2)--(-1.6,-2.3)--(-4.4,-2.3);
\draw  (1.5,-1.5) node {Prolegs};
\draw  (2,2) node {Spiracles};
\draw  (-1,3) node {Third abdominal};
\draw  (-1,2.5) node {segment};
\draw  (-3,-1.5) node {Ventral internal};
\draw  (-3,-2) node {lateral muscle};
\draw [line width=1.0pt,black](1.5,-1.2) -- (2.5,-0.2);
\draw [line width=1.0pt,black](1.5,-1.2) -- (1.3,-0.5);
\draw [line width=1.0pt,black](2,1.7) -- (0.9,0.2);
\draw [line width=1.0pt,black](2,1.7) -- (3.1,0.9);
\draw [line width=1.0pt,black](-1,0.2) -- (-1,2.3);
\filldraw[fill=black!90,fill opacity=0.4,draw=black!90,draw opacity=0.4](-0.6,-0.45)--(-0.55,-0.8)--(-1.45,-0.8)--(-1.4,-0.45)--(-0.6,-0.45);
\draw [line width=1.0pt,black](-1,-0.6) -- (-3,-1.2);
\end{tikzpicture} 
\caption{A fifth instar \textit{Manduca sexta} caterpillar.}
\label{fig:7}
\end{figure}

A fifth instar larva is a caterpillar in the final stage of  growth before transformation into a pupa.   Woods et al.\@
 \cite{woods} described the muscles of a fifth instar larva as being typically 4-6mm long and consisting of 2-14 fibres with a typical \textit{Manduca sexta}  caterpillar having approximately 70 muscles per larval segment. Nearly all the muscles are oriented longitudinally or obliquely and not circumferentially. The ventral interior longitudinal muscle is a comparatively large muscle and lies in the inner layer of  muscle, approximately spanning the ends of each proleg and expanding laterally to the spiracles, as shown in Figure \ref{fig:7}. 

 Dorfmann et al.\@ \cite{dorfmann2007} compared the  experimental data of carbon filled rubber vulcanizates for uniaxial cyclic stretching with the   experimental data of a ventral internal lateral caterpillar muscle, taken from the third abdominal body segment of a fifth instar larva.   They observed that the mechanical response of the two materials was similar,  both materials exhibiting stress-relaxation, hysteresis, residual strain and creep of residual strain, even though they differed in strength and stiffness.

The ventral internal lateral caterpillar muscle is considered to be a non-linear pseudo-elastic composite having a single preferred direction and so it can be regarded as being transversely isotropic. Incompressibility is assumed since changes in volume under deformation within the physiological range are very  small and therefore may be regarded as negligible.

If we assume that the biological material is, and remains, homogeneous throughout the deformation, then the material may be modelled using the constitutive equation (\ref{eq:7.2w}). 

For an isochoric uniaxial deformation in which $\lambda_1=\lambda>1$ and $ \lambda_2 =\lambda_3 =\lambda^{- 1/2}$ the  right Cauchy-Green strain tensor is
\begin{displaymath}
\textbf{C}= \left( \begin{array}{ccc}	
{\lambda^2}& 0 &0\\
0 &{\lambda^{-1}}& 0\\
0& 0 &{\lambda^{-1}} \end{array}\right).
\end{displaymath}

Biological tissue will soften in the direction of increasing strain, i.e. the 1-direction. Throughout this paper, for transversely isotropic deformations we assume that the 1-direction is aligned with the preferred inherent material direction. The preferred direction is then a unit vector $\mathbf{A}$ in the 1-direction, given by
\[ \mathbf{A} = \left(\begin{array}{c}1\\0\\0\end{array}\right). \]

The five invariants of incompressible transverse isotropy in this uniaxial extension are 
\begin{align*}
I_1 &= \lambda^2+2\lambda^{-1},\quad I_2 = \lambda^{-2}+2\lambda, \quad I_3=1,\\
I_4 &= \lambda^2,\quad I_5 = \lambda^4.
\end{align*}

To obtain from (\ref{eq:7.2w}) the single non-zero stress component  $T^{\ti}_{11}$ of a uniaxial tension, we may fix the value of $p$ by the requirement
$
\textbf{T}^{\mathscr{E}_{\iso}}_{22}(\lambda)=\textbf{T}^{\mathscr{E}_{\iso}}_{33}(\lambda)=0
$
because of the vanishing of the lateral stresses. 

In order to facilitate comparison between theory and experiment in modelling the Mullins effect we have used the engineering (nominal) stress component $T_{E11}$ defined by
\[
T_{E11}=\lambda^{-1}T_{11}.
\]
In the present situation of a diagonal deformation gradient and Cauchy stress the engineering stress employed here is the same as the Biot stress of  Rickaby \& Scott \cite{rickaby1, rickaby}. 

\begin{figure}[ht]
\centerline{
\includegraphics[width=18cm,height=10cm]{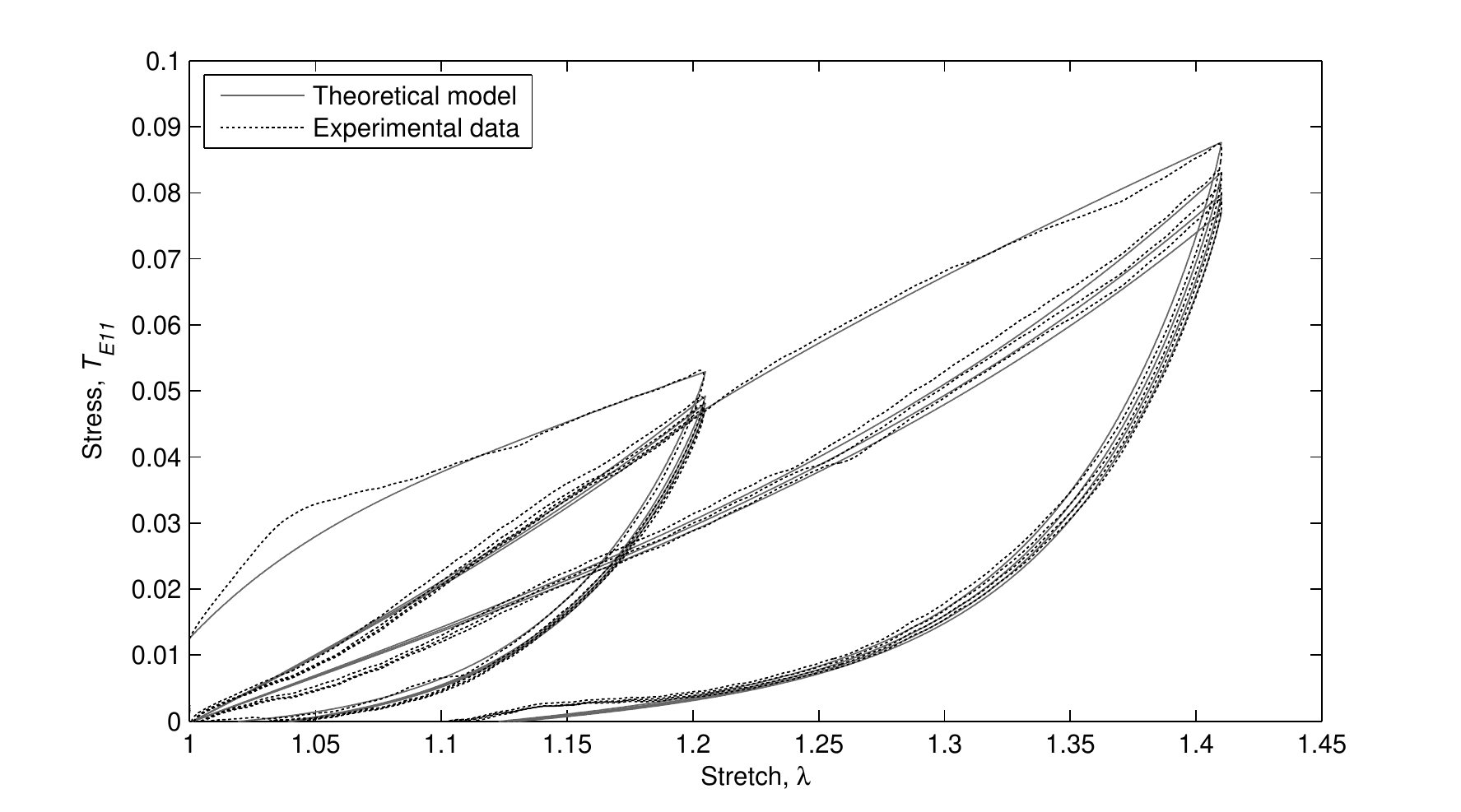}}
\caption{Comparison of our theoretical model with experimental data of  Dorfmann et al.\@ \cite{dorfmann2007}.}
\label{fig:8}
\end{figure}

Figure \ref{fig:8} provides a comparison between the softening model developed here and   experimental data for the caterpillar muscle. The experimental data came courtesy of  Dorfmann et al.\@  \cite{dorfmann2007} and were presented in their paper. In our theoretical model we employ the Cohen \cite{cohen} approximation
\[ \mathscr{L}^{-1}(x) \approx  \frac{3x-x^3}{1-x^2}\]
and the Heaviside step function $H(t)$   defined by
\[
H(t)=\left\{ \begin{array}{clrr}
0& t<0,\\
1& t\geq 0.
\end{array}\right.
\]
For comparison with  Dorfmann et al.\@  \cite{dorfmann2007} we adopt the following constants and functions:
\[
N=2.5, \quad \mu=0.0532, \quad r_{1,2}=1.06, \quad \alpha^2=0.4,\quad  a(t)=H(t-t_1)(t-t_1), \quad a_1=0.4,
\]
\[
{A}_0=-0.008, \quad A_{1,2}(t)=-0.008\log(\phi_{n,\omega}t), \quad A_{4,5}(t)=0, 
\]
\[
 \zeta_0(\lambda)=1-[{\tanhs(\lambda_{\C\_1}-\lambda)}]^{2.7},
\]
\[
\vartheta_\omega=\left\{\begin{array}{clrr}
0.83\\
1.00\\
\end{array}\right.
\quad
s_1=\left\{ \begin{array}{clrr}
0.246\\
0.120\\
\end{array}\right.
\quad
s_2=\left\{\begin{array}{clrr}
-0.90& \textrm{unloading},\\
-0.04& \textrm{loading}.\\
\end{array}\right.
\]
For $\lambda_{\C\_1}=1.205$
\[
d_\omega(\lambda_{\C\_1})=\left\{ \begin{array}{clrr}
0.00002 \,\lambda_{\C\_1}\\
0.0 \phantom{0000\,\lambda_{\C\_1}}\\
\end{array}\right.
\quad
\phi_{1,\omega}=\left\{\begin{array}{clrr}
0.5\\
0.5\\
\end{array}\right.
\quad
\mu b_\omega=\left\{ \begin{array}{clrr}
0.23& \textrm{unloading},\\
0.65& \textrm{loading}.\\
\end{array}\right.
\]
For $\lambda_{\C\_2}=1.410$
\[
d_\omega(\lambda_{\C\_2})=\left\{ \begin{array}{clrr}
0.0002 \,\lambda_{\C\_2}\\
0.0 \phantom{000\,\lambda_{\C\_2}}\\ \end{array}\right.
\quad
\phi_{2,\omega}=\left\{\begin{array}{clrr}
2.0\\
2.0\\
\end{array}\right.
\quad
\mu b_\omega=\left\{ \begin{array}{clrr}
0.33& \textrm{unloading},\\
0.80& \textrm{loading}.\\
\end{array}\right.
\]

It can be seen from Figure \ref{fig:8} that our model compares very well with the experimental data.

\subsection{Human thoracic aorta}
\begin{figure}[ht]
\centering
\begin{tikzpicture}[scale=1.15]
[decoration={random steps,segment length=4pt,amplitude=1.1pt}]
\filldraw[rotate=-7, fill=black!30, draw=black,decorate] (9.05,2.35) arc (-50:-2:80pt and 35pt)-- (9.6,7.15)--(8.2,7.20);
\filldraw[rotate=-7,fill=black!50,draw=black,decorate](8.6,6.3) arc (-50:183:80pt and 35pt);
\filldraw[rotate=-7,fill=black!30,draw=black](8.17,6.44) arc (-48:172:61.2pt and 26pt)--(4.5,6.3);
\filldraw[rotate=-7,fill=black!10,draw=black](8.17,6.44) arc (-48:-20:61.2pt and 26pt)--(8.73,6.85) arc (20.5:168:61pt and 26pt)--(4.825,3.65);
\filldraw[rotate=-7,draw=black,pattern=crosshatch](8.17,6.44) arc (-48:-20:61.2pt and 26pt)--(8.73,6.85) arc (20.5:168:61pt and 26pt)--(4.825,3.65);
\filldraw[rotate=-7,fill=black!0,draw=black](8.17,6.44) arc (-48:-20:61.2pt and 26pt)--(8.7,6.8) arc (22.5:185:61pt and 26pt)--(4.825,3.65);
\filldraw[rotate=-7,draw=black,pattern=crosshatch](8.17,6.44) arc (-48:-20:61.2pt and 26pt)--(8.7,6.8) arc (22.5:185:61pt and 26pt)--(4.825,3.65);
\filldraw[rotate=-7,fill=black!60,draw=black](8.17,6.44) arc (-48:-45:61.2pt and 26pt)--(8.34,6.54) arc (38.5:165:59pt and 25pt);
\filldraw[rotate=-7,fill=black!40,draw=black](8.5,5.47)--(8.15,6.47) arc (42:152:53.5pt and 21pt)--(5.0,3.65);
\filldraw[rotate=-7,fill=black!30,draw=black](8.5,5.47)--(8.35,5.95) arc (30:158:53.5pt and 21pt);
\filldraw[rotate=-7,fill=black!10,draw=black](8.17,5.64)--(8.16,5.98) arc (42:148:53.5pt and 21pt)--(5.1,3.65);
\filldraw[rotate=-7,fill=black!60,draw=black](8.17,5.64)--(8.16,5.71) arc (42:148:53.5pt and 24pt);
\filldraw[rotate=-7,fill=black!40,draw=black](8.6,3.5)--(8.17,5.64) arc (32:138:46pt and 17pt)--(5.1,3.65);
\filldraw[rotate=-7,fill=black!60,draw=black](8.6,4.5)--(7.98,5.52) arc (-50:-41:51.0pt and 21pt);
\filldraw[rotate=-7,fill=black!10,draw=black](8.6,4.5)--(8.17,5.24) arc (32:138:45pt and 17pt)--(5.1,3.65);
\filldraw[rotate=-7,fill=black!00,draw=black](8.6,4.5)--(8.17,5.14) arc (32:138:46pt and 17pt)--(5.1,3.65);
\filldraw[rotate=-7,draw=black,pattern=crosshatch dots](8.6,4.5)--(8.17,5.24) arc (32:138:45pt and 17pt)--(5.1,3.65);
\filldraw[rotate=-7,draw=black,pattern=crosshatch dots](8.6,4.5)--(8.17,5.14) arc (32:138:46pt and 17pt)--(5.1,3.65);
\filldraw[rotate=-7,fill=black!30,draw=black](8.6,3.75)--(7.88,4.85) arc (-53:-45:47pt and 17pt);
\filldraw[rotate=-7,fill=black!30,draw=black](8.35,4.95)--(7.94,5.06) arc (-50:-40:53.5pt and 21pt);
\filldraw[rotate=-7,fill=black!50,draw=black](7.88,4.74) arc (49:131:47pt and 17pt)--(8.5,3);
\filldraw[rotate=-7,fill=black!30,draw=black](8.5,3.5)--(7.88,4.68) arc (43:115:41pt and 14pt)--(6.15,3.6);
\filldraw[rotate=-7,fill=black!30,draw=black](8.5,3.7)--(7.75,3.42) arc (58:114:42pt and 15pt)--(6.15,3.7);
\filldraw[rotate=-7,fill=black!50,draw=black] (7.68,3.91) arc (-50:-35:47pt and 17pt)--(8.4,2.85);
\filldraw[rotate=-7,fill=black!10,draw=black](8.5,5.2)--(8.13,5.94) arc (-50:-30:60pt and 26pt);
\filldraw[rotate=-7,draw=black,pattern=crosshatch](8.5,5.2)--(8.13,5.94) arc (-50:-30:60pt and 26pt);
\draw[rotate=-7,draw=black](7.84,4.24) arc (-50:-40:46pt and 17pt);
\filldraw[fill=black!40,draw=black](6.2,2.65)--(6.75,2.65)--(6.75,4.0)--(6.15,4.0);
\filldraw[fill=black!0,draw=black](6.2,2.65)--(6.35,2.65)--(6.35,4.55)--(6.2,4.55);
\filldraw[draw=black,pattern=crosshatch dots](6.2,2.65)--(6.35,2.65)--(6.35,4.55)--(6.2,4.55);
\filldraw[fill=black!40, draw=black](5.8,2.65) -- (6.25,2.65)--(6.25,5.0)--(5.8,5.0);
\filldraw[fill=black!20, draw=black](5.7,2.65) -- (6.05,2.65)--(6.05,5.3)--(5.7,5.3);
\filldraw[fill=black!40, draw=black](8.45,1.6)--(8.1,1.75)-- (8.1,2.95) --(8.45,2.85);
\filldraw[fill=black!0, draw=black](8.45,1.62)--(8.30,1.65)-- (8.30,3.25) --(8.45,3.25);
\filldraw[draw=black,pattern=crosshatch dots](8.45,1.62)--(8.30,1.65)-- (8.30,3.25) --(8.45,3.25);
\filldraw[fill=black!40, draw=black](8.65,1.5)--(8.4,1.6)--(8.4,3.85)--(8.65,3.75);
\filldraw[fill=black!20, draw=black](8.65,1.5)--(8.5,1.55)--(8.5,4.05)--(8.65,3.95);
\filldraw[fill=black!50, draw=black] (8.9,1.4)--(8.6,1.5)--(8.6,4.5)--(8.9,4.4);
\filldraw[fill=black!50, draw=black] (5.5,2.65) --(5.825,2.65) -- (5.825,5.65)--(5.25,5.65);
\filldraw[fill=black!0, draw=black](5.36,6.15)--(5.50,6.15)--(5.50,2.65)--(5.36,2.65);
\filldraw[fill=black!0, draw=black](8.9,1.4)--(8.8,1.45)--(8.8,4.9)--(8.9,4.85);
\filldraw[draw=black,pattern=crosshatch](5.36,6.15)--(5.50,6.15)--(5.50,2.65)--(5.36,2.65);
\filldraw[draw=black,pattern=crosshatch](8.9,1.4)--(8.8,1.45)--(8.8,4.9)--(8.9,4.85);
\filldraw[fill=black!40, draw=black,decorate] (9.25,5.22)-- (9.3,5.20)--(9.3,1.22)--(9.25,1.24);
\filldraw[fill=black!40, draw=black] (9.25,1.24)--(8.9,1.4) -- (8.9,5.4)--(9.25,5.22);
\filldraw[fill=black!40, draw=black,decorate] (4.9,2.65)--(4.825,2.65) -- (4.825,6.65) -- (4.9,6.65);
\filldraw[fill=black!40, draw=black](4.9,6.65) --(5.45,6.65) -- (5.45,2.65)--(4.9,2.65);
\coordinate [label=right:{Tunica intima}] () at (0.5,0.5);
\coordinate [label=right:{Tunica media}] () at (0.5,1);
\coordinate [label=right:{Tunica adventitia}] () at (0.5,1.5);
\draw [line width=1.0pt,black](4.8,2.5) -- (5.4,2.5);
\draw [line width=1.0pt,black](5.15,1.5) -- (5.15,2.5);
\draw [line width=1.0pt,black](5.5,2.5) -- (6.2,2.5);
\draw [line width=1.0pt,black](5.85,1) -- (5.85,2.5);
\draw [line width=1.0pt,black](6.3,2.5) -- (6.8,2.5);
\draw [line width=1.0pt,black](6.55,0.5) -- (6.55,2.5);
\draw [line width=1.0pt,black](4,0.5) -- (6.55,0.5);
\draw [line width=1.0pt,black](4,1) -- (5.85,1);
\draw [line width=1.0pt,black](4,1.5) -- (5.15,1.5);
\coordinate [label=right:{Internal elastic lamina}] () at (11,2);
\draw [->,line width=1.0pt,black](10.8,2) -- (7.5,4.3);
\coordinate [label=right:{Fibrocollagenous tissue}] () at (11,5);
\draw [->,line width=1.0pt,black](10.8,5) -- (7.5,5.6);
\coordinate [label=right:{Smooth muscle}] () at (11,4);
\draw [->,line width=1.0pt,black](10.8,4) -- (7.5,5.25);
\coordinate [label=right:{Fibrocollagenous tissue}] () at (11,3);
\draw [->,line width=1.0pt,black](10.8,3) -- (7.5,4.9);
\coordinate [label=right:{Fibrocollagenous tissue}] () at (11,7);
\draw [->,line width=1.0pt,black](10.8,7) -- (7.5,6.9);
\coordinate [label=right:{External elastic lamina}] () at (11,6);
\draw [->,line width=1.0pt,black](10.8,6) -- (7.5,6.2);
\coordinate [label=right:{Endothelium}] () at (11,1);
\draw [->,line width=1.0pt,black](10.8,1) -- (7.5,3);
\end{tikzpicture}
\caption{Representation of the histology of an elastic artery.}
\label{fig:9}
\end{figure}
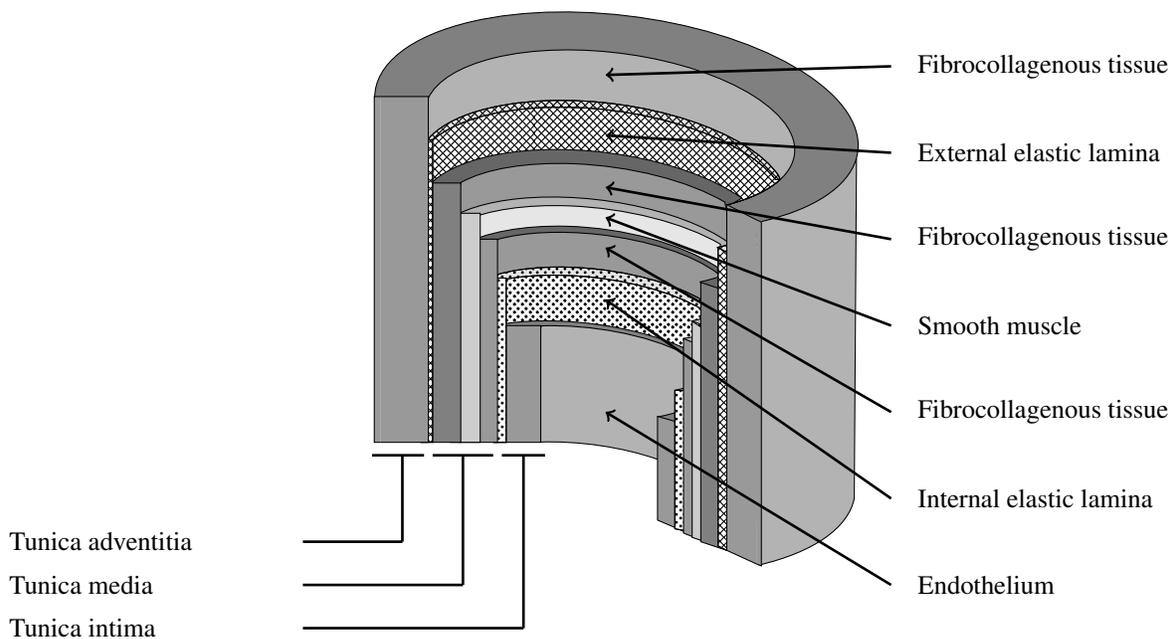

The histology of an elastic artery is represented in Figure \ref{fig:9}. All arteries are composed of three layers, the tunica intima, tunica media and tunica adventitia. The tunica intima is the innermost layer consisting of an elastic membrane and a monolayer of smooth endothelial cells. The tunica media is the middle layer which is comprised of smooth muscle, elastin and bundles of collagen fibres. The tunica adventitia is the outermost layer which consists of connective tissues, collagen and elastic fibres.  Humphrey \cite[page 256]{humphrey} comments that the orientation and distribution of the medial constituents vary with species and location along the vascular tree. For different species Gasser et al.\@ \cite{gasser} observed  the following similarities: the medial collagen fibres, elastin and smooth muscle cells tend to be almost circumferentially oriented, the advential collagen fibres tend to have an axial orientation, whilst the intima elastic membrane and endothelial cells have a non-uniform orientation.

Further discussion on the histology of the human artery can be found in either Humphrey \cite[pages 254--259]{humphrey} or  Gasser et al.\@ \cite{gasser} and the works cited  therein.

\begin{figure}[ht]
\centering
\begin{tikzpicture}[scale=0.9]
\draw (2.5,1.5) circle (1.5cm);
\draw (2.5,1.5) circle (1.2cm);
\draw [dashed] (11.10,5.2) arc (293:130:1.5cm);
\draw  (11.10,5.2) arc (-67.0:135:1.5cm);
\draw [dashed] (10.97,5.42) arc (293:130:1.2cm);
\draw  [dashed] (10.97,5.42) arc (-67:130:1.2cm);
\draw (3.1,0.11) -- (11.10,5.2);
\draw  (1.9,2.89) -- (9.7,7.85);
\draw [black!30] (4.9,1.57) arc (-22:90:1.5cm);
\draw [black!30] (5.9,2.2) arc (-22:90:1.5cm);
\draw [black!30,dashed] (4.6,1.57) arc (-28:90:1.2cm);
\draw [black!30,dashed] (5.6,2.2) arc (-28:90:1.2cm);
\draw [black!30] (5.9,2.2) -- (4.9,1.57);
\draw [black!30] (4.5,4.275) -- (3.5,3.645);
\draw [black!30,dashed] (5.6,2.2) -- (4.6,1.57);
\draw [black!30,dashed] (4.5,3.975) -- (3.5,3.345);
\draw [black!30,dashed] (5.6,2.2)--(5.9,2.2);
\draw [black!30,dashed] (4.6,1.57)--(4.9,1.57);
\draw [black!30,dashed] (4.5,3.975)--(4.5,4.275);
\draw [black!30,dashed] (3.5,3.345)--(3.5,3.645);
\draw [black!30] (9,4.68) arc (0:60:1.5cm);
\draw [black!30] (7.0,3.41) arc (0:60:1.5cm);
\draw [black!30,dashed] (8.7,4.68) arc (-2:53:1.2cm);
\draw [black!30,dashed] (6.7,3.41) arc (-2:53:1.2cm);
\draw [black!30] (8.25,5.98) -- (6.25,4.71);
\draw [black!30] (9,4.68) -- (7.0,3.41);
\draw [black!30,dashed] (8.25,5.68) -- (6.25,4.41);
\draw [black!30,dashed] (8.7,4.68) -- (6.7,3.41);
\draw [black!30,dashed] (8.25,5.68) -- (8.25,5.98);
\draw [black!30,dashed] (6.25,4.41) -- (6.25,4.71);
\draw [black!30,dashed] (8.7,4.68) -- (9,4.68);
\draw [black!30,dashed] (6.7,3.41) -- (7.0,3.41);
\draw [dashed] (0.5,0.23) -- (11,6.91);
\draw [black!30, dashed] (11,1)--(14,1)--(14,2.5)--(11,2.5)--(11,1);
\draw [black!30] (10.8,0.8)--(13.8,0.8)--(13.8,2.3)--(10.8,2.3)--(10.8,0.8);
\draw [black!30,dashed] (11,1)--(10.8,0.8);
\draw [black!30,dashed] (14,1)--(13.8,0.8);
\draw [black!30,dashed] (14,2.5)--(13.8,2.3);
\draw [black!30,dashed] (11,2.5)--(10.8,2.3);
\draw [->] (12.3,1.55)--(14.8,1.55);
\draw [->] (12.3,1.55)--(12.3,3.3);
\draw [->] (12.3,1.55)--(10.9,0.25);
\coordinate [label=right:{$\lambda_1$}] (C) at (14.8,1.55);
\coordinate [label=above:{$\lambda_2$}] (C) at (12.3,3.3);
\coordinate [label=below:{$\lambda_3$}] (C) at (10.8,0.35);
\coordinate [label=above:{$A$}] (A) at (4.6,3.3);
\coordinate [label=above:{$B$}] (B) at (7.75,4.2);
\end{tikzpicture}
\caption{Samples taken from the thoracic aorta.}
\label{fig:10}
\end{figure}
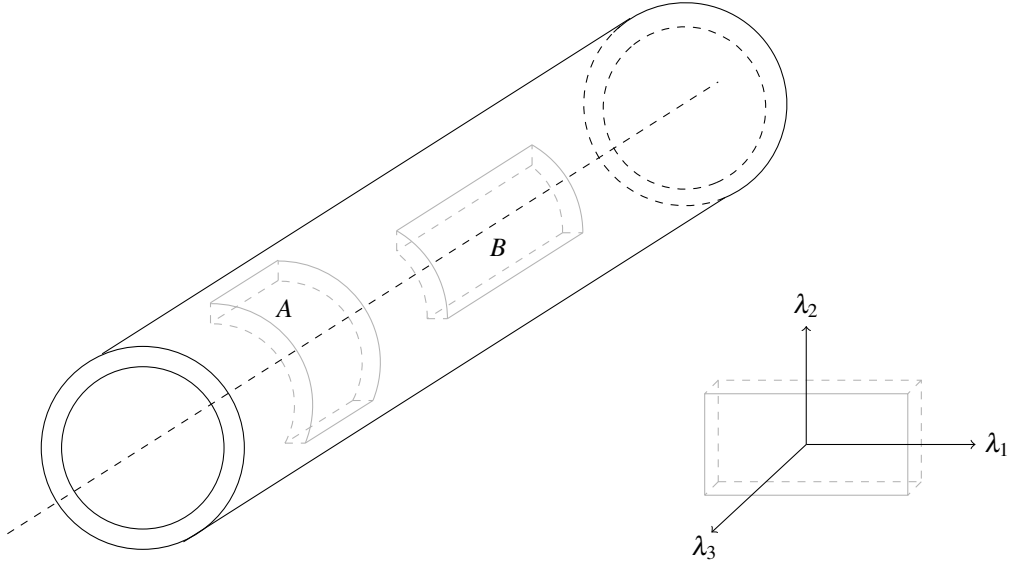

 Horn\'y et al.\@ \cite{horny} conducted cyclic uniaxial tests on two samples of  human thoracic aorta, which is the largest artery in the body, the first being  sample $A$ of Figure \ref{fig:10} taken in the circumferential direction,  and the second being sample $B$ of Figure \ref{fig:10} taken in the longitudinal direction. 

 Horn\'y et al.\@  \cite{horny} considered that the aorta could be modelled as  an anisotropic, incompressible  hyperelastic material. The incompressibility assumption is substantiated by Carew et al.\@  \cite{carew} who conducted tests on samples of canine thoracic aortas, concluding that during compression tests within the physiological range there was minimal volume change in the thoracic aorta samples.  Clark \&  Glagov \cite{clark} studied the histology of arterial material and showed that arteries are composed of reinforced fibre layers. Humphrey \cite[pages 267--268]{humphrey} observed that each separate layer is homogeneous with the mechanical properties varying between the layers. We therefore assume that the arterial material is, and remains, homogeneous throughout the deformation.  As a result of the inherent layering of the media, depicted in Figure \ref{fig:9}, we assume that the orthotropic artery may be represented by two transversely isotropic layers, one for the preferred circumferential direction and the other for the preferred longitudinal direction. This approach is used by several authors including Holzapfel et al.\@ \cite{holzapfel2000}.  For these reasons we consider that the biological material may be modelled using the transversely isotropic constitutive equation (\ref{eq:7.2w}).

\begin{figure}[ht]
\centerline{
\includegraphics[width=18cm,height=10cm]{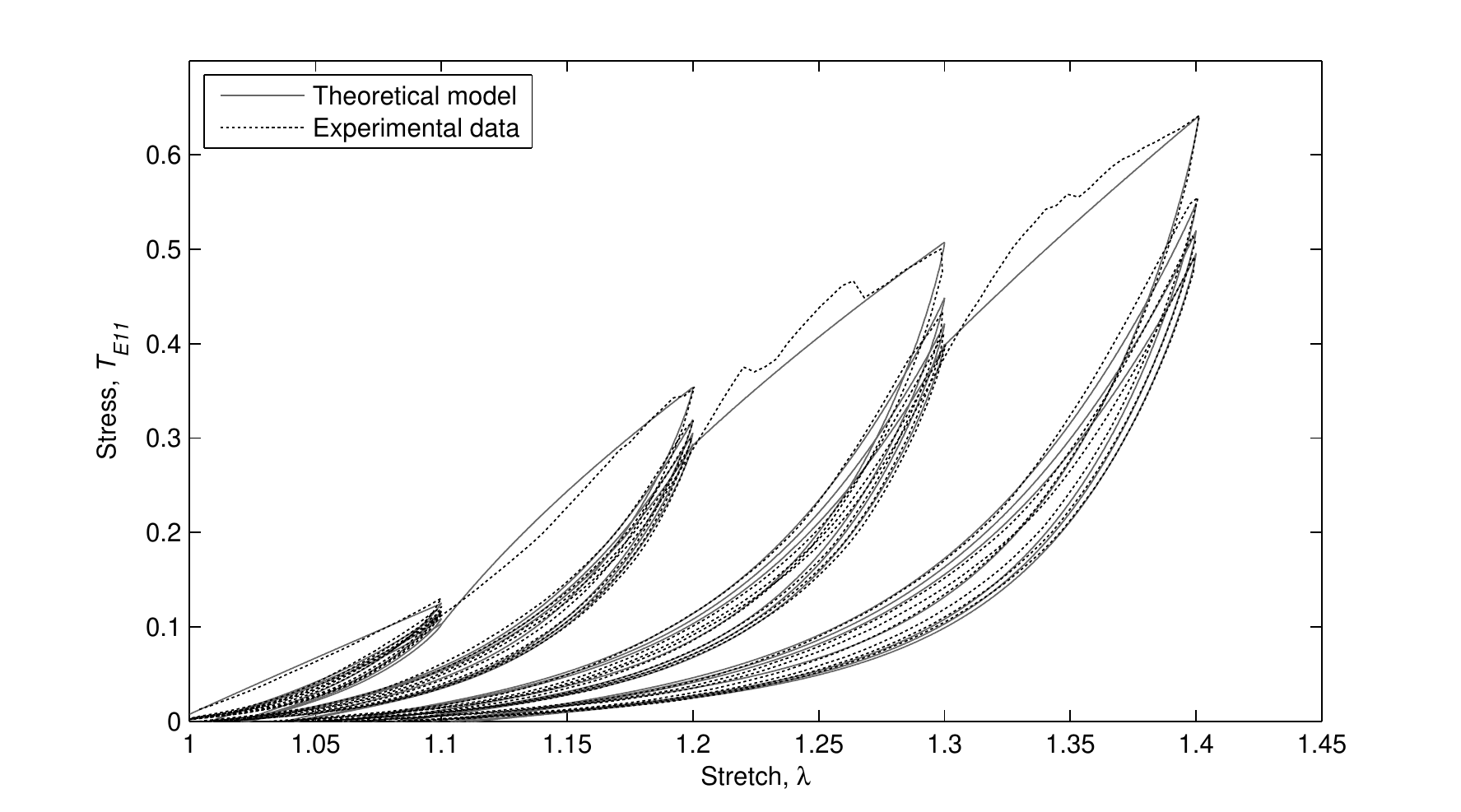}}
\caption{Comparison of our theoretical model with experimental data of Horn\'y et al.\@ \cite{horny}, for the uniaxial circumferential deformation}
\label{fig:11}
\end{figure}

Figure \ref{fig:11} provides a comparison between the  softening model developed here and the   experimental data for a uniaxial circumferential deformation. This experimental data came courtesy of Horn\'y et al.\@ \cite{horny} and was presented in their paper.  In our model we employ the following constants and functions:
\[
N=10, \quad \mu=0.425, \quad r_\omega=1.06, \quad \alpha^2=0.4, \quad  a(t)=H(t-t_1)(t-t_1), \quad a_1=0.4,
\]
\[
{A}_0=-0.5, \quad A_{1,2}(t)=-0.1\log(\phi_{n,\omega}t), \quad A_{4,5}(t)=0,
\]
\[
\zeta_0(\lambda)=1-{\tanhs(\lambda_{\C\_1}-\lambda)},
\]
\[
\vartheta_\omega=\left\{\begin{array}{clrr}
0.80\\
0.94\\
\end{array}\right.
\quad
s_1=\left\{ \begin{array}{clrr}
0.612\\
0.612\\
\end{array}\right.
\quad
s_2=\left\{\begin{array}{clrr}
0.08& \textrm{unloading},\\
0.08& \textrm{loading}.\\
\end{array}\right.
\]
\noindent{
For $\lambda_{\C\_1}=1.1$}
\[
d_\omega(\lambda_{\C\_1})=\left\{ \begin{array}{clrr}
0.00013 \,\lambda_{\C\_1}\\
0.00004 \,\lambda_{\C\_1}\\ \end{array}\right.
\quad
\phi_{1,\omega}=\left\{\begin{array}{clrr}
5.0\\
5.0\\
\end{array}\right.
\quad
\mu b_\omega=\left\{ \begin{array}{clrr}
0.09& \textrm{unloading},\\
0.08& \textrm{loading}.\\
\end{array}\right.
\quad
\]
For $\lambda_{\C\_2}=1.2$
\[
d_\omega(\lambda_{\C\_2})=\left\{ \begin{array}{clrr}
0.0006 \,\lambda_{\C\_2}\\
0.0002 \,\lambda_{\C\_2}\\ \end{array}\right.
\quad
\phi_{2,\omega}=\left\{\begin{array}{clrr}
0.5\\
0.5\\
\end{array}\right.
\quad
\mu b_\omega=\left\{ \begin{array}{clrr}
0.11& \textrm{unloading},\\
0.08& \textrm{loading}.\\
\end{array}\right.
\quad
\]
For $\lambda_{\C\_3}=1.3$
\[
d_\omega(\lambda_{\C\_3})=\left\{ \begin{array}{clrr}
0.0014 \,\lambda_{\C\_3}\\
0.0006 \,\lambda_{\C\_3}\\ \end{array}\right.
\quad
\phi_{3,\omega}=\left\{\begin{array}{clrr}
0.5\\
0.5\\
\end{array}\right.
\quad
\mu b_\omega=\left\{ \begin{array}{clrr}
0.13& \textrm{unloading},\\
0.11& \textrm{loading}.\\
\end{array}\right.
\quad
\]
For $\lambda_{\C\_4}=1.4$
\[
d_\omega(\lambda_{\C\_4})=\left\{ \begin{array}{clrr}
0.005 \,\lambda_{\C\_4}\\
0.004 \,\lambda_{\C\_4}\\ \end{array}\right.
\quad
\phi_{4,\omega}=\left\{\begin{array}{clrr}
0.5\\
0.5\\
\end{array}\right.
\quad
\mu b_\omega=\left\{ \begin{array}{clrr}
0.14& \textrm{unloading},\\
0.14& \textrm{loading}.\\
\end{array}\right.
\quad
\]

We may read off from the graphs presented in Figure \ref{fig:11}  that at the start of unloading for cycle 1 the stress is approximately $0.125$ MPa, at the start of unloading for cycle 2 the stress has increased by approximately  $0.229$ MPa, at the start of unloading for cycle 3 the stress has increased by approximately $0.149$ MPa, whilst at the start of unloading for cycle 4 the stress has increased by approximately $0.134$~MPa.  The increase in stress at the start of unloading for cycle 2 is therefore significantly greater than the stress increases at the start of unloading for the other cycles 1, 3, 4, which are all similar.  The raised stress for cycle 2  has been reflected in the model by choosing   constants  $\phi_{2,\omega}$ different from those in the other cycles.

The stress softening on the new primary loading paths of cycles 3 and 4 is under-predicted by the present model, a feature which we believe may be corrected by having non-monotoic functions $A_l(t)$ defined at equation (\ref{eq:5.3w}).

\begin{figure}
\centerline{
\includegraphics[width=18cm,height=10cm]{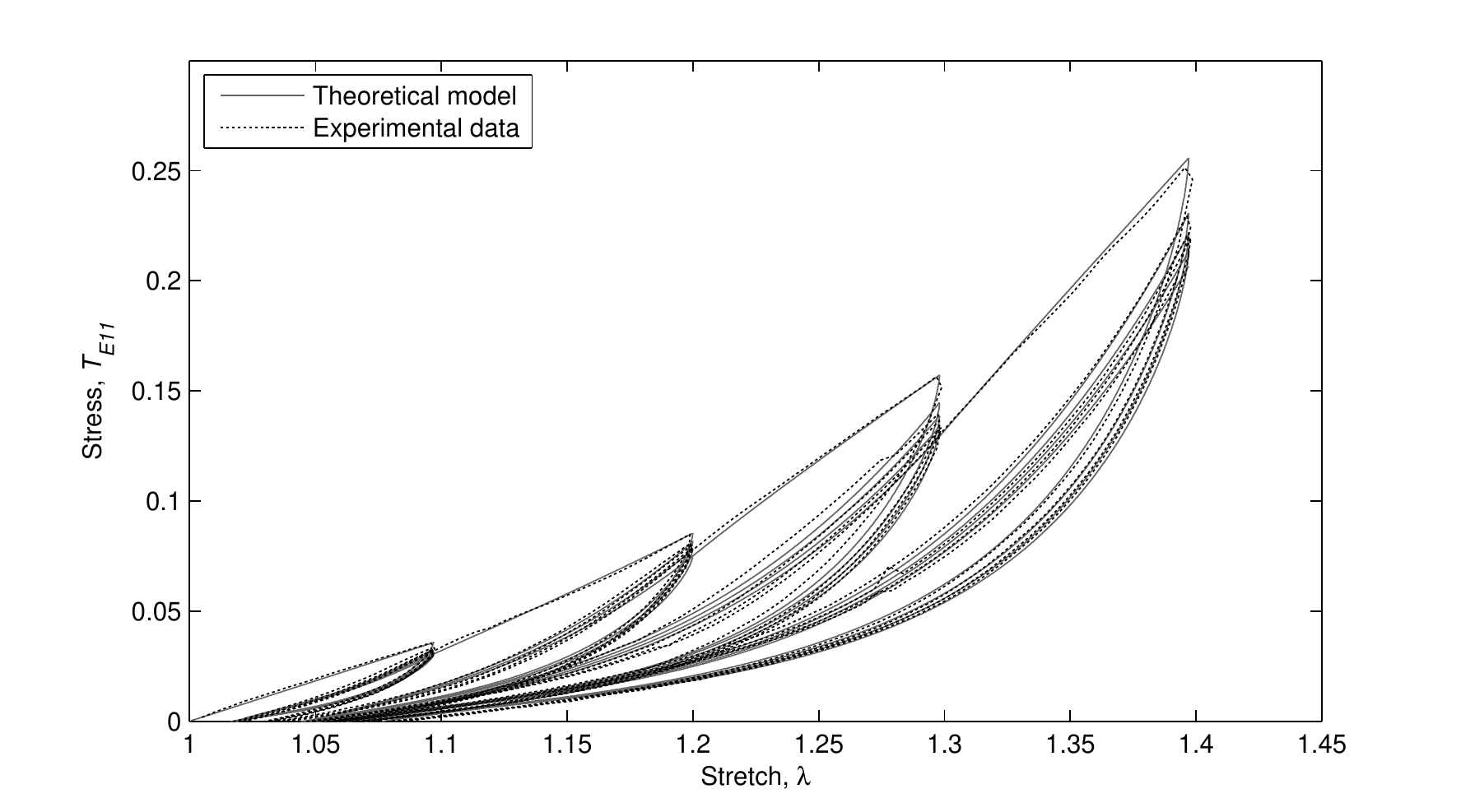}}
\caption{Comparison of our theoretical model with experimental data of 
Horn\'y et al.\@ \cite{horny}, for the uniaxial longitudinal deformation.}
\label{fig:12}
\end{figure}

Figure \ref{fig:12} provides a comparison between the softening model developed here and the   experimental data for a uniaxial longitudinal deformation. This experimental data came courtesy of Horn\'y et al.\@ \cite{horny} and was presented in their paper. In our model we use  the following constants and functions:

\[
N=10, \quad \mu=0.125, \quad r_1=r_2=1.06, \quad \alpha^2=0.4, \quad  a(t)=H(t-t_1)(t-t_1), \quad a_1=0.4,
\]
\[
{A}_0=-0.2, \quad A_{1,2}(t)=-0.03\log(\phi_{n,\omega}t), \quad A_{4,5}(t)=0, 
\]
\[
\zeta_0(\lambda)=1-{\tanhs(\lambda_{\C\_1}-\lambda)},  
\]
\[
\vartheta_\omega=\left\{\begin{array}{clrr}
0.65\\
0.97\\
\end{array}\right.
\quad
s_1=\left\{\begin{array}{clrr}
0.24\\
0.24\\
\end{array}\right.
\quad
s_2=\left\{ \begin{array}{clrr}
-0.30& \textrm{unloading},\\
-0.30& \textrm{loading}.\\
\end{array}\right.
\]
For $\lambda_{\C\_1}=1.1$
\[
d_\omega(\lambda_{\C\_1})=\left\{ \begin{array}{clrr}
0.00004 \,\lambda_{\C\_1}\\
0.00004 \,\lambda_{\C\_1}\\ \end{array}\right.
\quad
\phi_{1,\omega}=\left\{\begin{array}{clrr}
0.13\\
0.13\\
\end{array}\right.
\quad
\mu b_\omega=\left\{ \begin{array}{clrr}
0.13& \textrm{unloading},\\
0.07& \textrm{loading}.\\
\end{array}\right.
\quad
\]
For $\lambda_{\C\_2}=1.2$
\[
d_\omega(\lambda_{\C\_2})=\left\{ \begin{array}{clrr}
0.0002 \,\lambda_{\C\_2}\\
0.0002 \,\lambda_{\C\_2}\\ \end{array}\right.
\quad
\phi_{2,\omega}=\left\{\begin{array}{clrr}
0.13\\
0.13\\
\end{array}\right.
\quad
\mu b_\omega=\left\{ \begin{array}{clrr}
0.18& \textrm{unloading},\\
0.11& \textrm{loading}.\\
\end{array}\right.
\quad
\]
For $\lambda_{\C\_3}=1.3$
\[
d_\omega(\lambda_{\C\_3})=\left\{ \begin{array}{clrr}
0.0004 \,\lambda_{\C\_3}\\
0.0004 \,\lambda_{\C\_3}\\ \end{array}\right.
\quad
\phi_{3,\omega}=\left\{\begin{array}{clrr}
0.05\\
0.05\\
\end{array}\right.
\quad
\mu b_\omega=\left\{ \begin{array}{clrr}
0.21& \textrm{unloading},\\
0.15& \textrm{loading}.\\
\end{array}\right.
\quad
\]
For $\lambda_{\C\_4}=1.4$
\[
d_\omega(\lambda_{\C\_4})=\left\{ \begin{array}{clrr}
0.0006 \,\lambda_{\C\_4}\\
0.0004 \,\lambda_{\C\_4}\\ \end{array}\right.
\quad
\phi_{4,\omega}=\left\{\begin{array}{clrr}
0.05\\
0.05\\
\end{array}\right.
\quad
\mu b_\omega=\left\{ \begin{array}{clrr}
0.24& \textrm{unloading},\\
0.16& \textrm{loading}.\\
\end{array}\right.
\quad
\]

We may read off from the graphs presented in Figure \ref{fig:12}  that at the start of unloading for cycle 1 the stress is approximately $0.036$ MPa, at the start of unloading for cycle 2 the stress has increased by approximately  $0.049$ MPa, at the start of unloading for cycle 3 the stress has increased by approximately $0.072$ MPa, whilst at the start of unloading for cycle 4 the stress has increased by approximately $0.098$~MPa.   The stress increases at the start of unloading for cycles 1 and 2 are roughly comparable as are the stress increases at the start of unloading for cycles 3 and 4.  However, the total stress increase for cycles 3 and 4 is nearly twice as large as that for cycles 1 and 2 taken together. This is directly reflected in the choice of $\phi_{n,\omega}$ constant with one set of values for cycles 1 and 2 and another set for cycles 3 and~4.

\section{Conclusions} 
\label{sec:conclusion}
We see from Figures \ref{fig:8}, \ref{fig:11} and \ref{fig:12} that the  transversely isotropic model developed here fits    the experimental data of soft biological tissue extremely well.

The model presented here appears to be the first in which transversely isotropic stress relaxation and residual strain models have been combined with a transversely isotropic Arruda-Boyce eight-chain model to develop a constitutive relation that is capable of accurately describing softening during multicyclic stress-strain  loading for a non-preconditioned biological material.

The results presented in Figures  \ref{fig:8}, \ref{fig:11} and \ref{fig:12} are by no means the only solutions that this model is capable of giving. By neglecting some of the inelastic terms, e.g. creep of residual strain, or selecting a single relaxation curve, we arrive at a simplified model with a reduced set of parameters. A similar parameter reduction could be obtained by using a preconditioned material such at that presented by  Dorfmann et al.\@   \cite[Figure 3]{dorfmann2008}. 

Once material parameters have been established for a specimen of biological tissue,  these parameters could be used to predict mulitcyclic softening of other biological materials with a similar molecular structure.  
 \textit{Manduca sexta} muscle is capable of undergoing large non-linear elastic deformations due to its shape-changing ability. As the \textit{Manduca sexta} muscle is
not attached to a jointed structure,  Dorfmann et al.\@  \cite{dorfmann2008} proposed that it would function like the muscles found in tongues, trunks and octopus arms, as described by Kier \& Smith \cite{kier}. Assuming these muscles are all of a similar type,  the softening model developed here could  represent all these muscles with a similar degree of accuracy. Following the work of Fung \cite{fung}, soft biological tissue such as the skin, ureter and arteries may also be represented by the  softening model developed here as these tissues also exhibit stress-relaxation, hysteresis and residual strain under cyclic loading and unloading conditions. 

For biaxial testing many authors consider that skin and other biological materials exhibit an  orthotropic stress-strain response, see Sacks \& Sun \cite{sacs2003} and the references cited therein. In order to model such materials we would need to develop an orthotropic version of the transversely isotropic model developed here.  We hope to develop such a model  in a subsequent paper.

\section*{Acknowledgements}
One of us (SRR) is grateful to  the University of East Anglia for the award of a PhD studentship. The authors thank Professor Luis Dorfmann and Dr Luk\'a\v s Horn\'y for most kindly supplying experimental data. We would also like to thank Mr Jeff Kraus for allowing us to publish the photograph of the \textit{Manduca sexta}. Furthermore, we would like to thank the reviewers for their helpful comments.

\bibliographystyle{plain}
\bibliography{BIBLIOGRAPHYa}

\end{document}